\documentclass[useAMS,usenatbib,iop,apj,tighten,11pt,a4paper]{emulateapj}

\usepackage{graphicx,color}
\usepackage{mathtools}	
\usepackage{bm}
\usepackage{ulem}
\usepackage{cases}

\def\aap{A\&A}
\def\apjl{ApJ}

\def\apjs{ApJS}

\renewcommand{\vec}[1]{{{\mbox{\boldmath $#1$}}}}

\newcommand{\Exp}[1]{{\rm e}^{#1}}

\newcommand{\del}{\partial}

\newcommand{\bfOmega}{\bm{\Omega}}

\newcommand{\mean}[1]{{#1}}
\newcommand{\meanv}[1]{\bm{#1}}

\newcommand{\eq}{_\mathrm{eq}}		
		
\newcommand{\f}{_\mathrm{0}}

\newcommand{\kin}{_\mathrm{k}}		
\newcommand{\magn}{_\mathrm{m}}

\newcommand{\crit}{_\mathrm{c}}		
\newcommand{\cor}{^\mathrm{c}}		
\newcommand{\sat}{_\mathrm{s}}		
\newcommand{\const}{\mathrm{const}}

\newcommand{\Co}{\mathrm{Co}}
\newcommand{\mbr}{\mean{B}_r}
\newcommand{\mbp}{\mean{B}_\phi}
\newcommand{\mbz}{\mean{B}_z}
\newcommand{\mbi}{\mean{B}_i}
\newcommand{\mur}{\mean{U}_r}
\newcommand{\mup}{\mean{U}_\phi}
\newcommand{\muz}{\mean{U}_z}

\newcommand{\muztilde}{\widetilde{\mean{U}}_z}

\newcommand{\arm}{_\mathrm{a}}

\newcommand{\Strouhal}{\mathrm{St}}
\newcommand{\Coriolis}{\mathrm{Co}}

\newcommand{\chisqred}{\chi^2_\nu}
\newcommand{\chisq}{\chi^2}
\newcommand{\Pearson}{_\mathrm{P}}
\newcommand{\adj}{_\mathrm{adj}}

\makeatletter
\newsavebox{\@brx}

\newcommand{\llangle}[1][]{\savebox{\@brx}{\(\m@th{#1\langle}\)}%
  \mathopen{\copy\@brx\kern-0.5\wd\@brx\usebox{\@brx}}}
\newcommand{\rrangle}[1][]{\savebox{\@brx}{\(\m@th{#1\rangle}\)}%
  \mathclose{\copy\@brx\kern-0.5\wd\@brx\usebox{\@brx}}}

  \newcommand{\cmcube}{\,{\rm cm^{-3}}}

  \newcommand{\kms}{\,{\rm km\,s^{-1}}}
  \newcommand{\kmskpc}{\,{\rm km\,s^{-1}\,kpc^{-1}}}

  \newcommand{\K}{\,{\rm K}}
  \newcommand{\kpc}{\,{\rm kpc}}
  \newcommand{\pc}{\,{\rm pc}}
  
  \newcommand{\Myr}{\,{\rm Myr}}
  
  \newcommand{\Gyr}{\,{\rm Gyr}}

  \newcommand{\mkG}{\,\mu{\rm G}}

\definecolor{webgreen}{rgb}{0,.5,0}
\definecolor{webbrown}{rgb}{.6,0,0}
\definecolor{purple}{rgb}{0.5,0,.5}

\begin{document}

\title{Statistical tests of galactic dynamo theory}
\author{Luke~Chamandy$^{1,2,3}$, Anvar~Shukurov$^{3}$ \&  A.~Russ~Taylor$^{1,2}$}
\affil{$^1$Astronomy Department, University of Cape Town, Rondebosch 7701, Republic of South Africa}
\affil{$^2$Department of Physics, University of the Western Cape, Belleville 7535, Republic of South Africa}
\affil{$^3$School of Mathematics \& Statistics, Newcastle University, Newcastle upon Tyne NE1 7RU}
\email{luke@ast.uct.ac.za}
\email{anvar.shukurov@ncl.ac.uk}
\email{russ@ast.uct.ac.za}

\begin{abstract}
Mean-field galactic dynamo theory is the leading theory to explain the prevalence of regular magnetic fields in spiral galaxies, 
but its systematic comparison with observations is still incomplete and fragmentary.
Here we compare predictions of mean-field dynamo models to observational data 
on magnetic pitch angle and the strength of the mean magnetic field.
We demonstrate that a standard $\alpha^2\Omega$ dynamo model
produces pitch angles of the regular magnetic fields of nearby galaxies 
that are reasonably consistent with available data. 
The dynamo estimates of the magnetic field strength are generally within a factor of a few of the observational values.
Reasonable agreement between theoretical and observed pitch angles generally requires 
the turbulent correlation time $\tau$ to be in the range $10$--$20\Myr$,
in agreement with standard estimates.
Moreover, good agreement also requires
that the ratio of the ionized gas scale height to root-mean-square turbulent velocity increases with radius.
Our results thus widen the possibilities to constrain interstellar medium (ISM) parameters
using observations of magnetic fields.
This work is a step toward systematic statistical tests of galactic dynamo theory.
Such studies are becoming more and more feasible as larger datasets are acquired using current and up-and-coming instruments.
\end{abstract}


\section{Introduction}
\label{sec:intro}
Spiral galaxies contain magnetic fields that are
coherent on scales larger than the outer scale of turbulence,
the so-called large-scale, or mean magnetic fields 
\citep{Beck+Wielebinski13,Beck15b}.
Mean-field turbulent dynamo theory is the leading theory to explain the prevalence and properties of 
large-scale magnetic fields in galaxies
\citep{Ruzmaikin+88,Beck+96,Brandenburg+Subramanian05a,Klein+Fletcher15}.
The galactic dynamo theory has proved to be successful in explaining the overall
large-scale magnetic properties of  
a generic spiral galaxy as well as those of a selection of nearby galaxies.
Dynamo models contain a number of parameters that are poorly constrained by theory and observations
as they require detailed knowledge on the size and shape of the galactic ionized layers, 
the magnitude and spatial distribution of the turbulent speeds and scales, as well
as their variations with galactic azimuth and radius
\citep[e.g.,][]{Brandenburg+93,SS98,Moss98,Rohde+99,Moss+07}.
Statistical studies of the large-scale magnetic fields in samples of galaxies are limited by 
the fact that they require observations of synchrotron emission and Faraday rotation
at sensitivity and resolution currently achievable only for a modest number of nearby galaxies
(\citealt{Fletcher10},
\defcitealias{Vaneck+15}{VBSF}\citealt[][(hereafter \citetalias{Vaneck+15})]{Vaneck+15}, 
\citealt{Tabatabaei+16}).

The radio telescopes of a new generation would significantly expand the observational database \citep[e.g.,][]{GBF04}
to make such comparisons feasible. 
However, approaches to such comparisons of theory with observations need to be developed now, 
but \citetalias{Vaneck+15} demonstrated that correlations
between individual parameters of spiral galaxies (such as the rotational shear rate) and their magnetic fields 
are not easy to detect because even the simpler properties
of the global magnetic structures are sensitive to a relatively large number of diverse galactic parameters. 
As a result, the scatter between the remaining parameters hides the expected correlation. In addition, the 
observational data need to be reduced in a coherent and systematic manner to admit comparison with theory.
Therefore, a statistical comparison of dynamo theory with observations requires, apart from a suitable reduction
of the observational data \citepalias[as suggested by][]{Vaneck+15}, a careful approach.

Another obstacle in observational verifications of galactic dynamo theory is that interstellar random magnetic fields
usually exceed the mean field in strength \citep{Ruzmaikin+88,Beck+96}, so that dedicated techniques are required to 
deduce the parameters of the global magnetic structures observed. Moreover, galactic magnetic fields, despite
being dominated by an axisymmetric structure (in agreement with predictions of the dynamo theory) 
are further modified by the spiral arms and various asymmetries of the host galaxies. Therefore, careful
identification of the underlying axially symmetric magnetic field in the observations is required before
the theory can be meaningfully compared with observations.  Among approaches suggested to quantify the large-scale
structures in the observed distributions of synchrotron emission and Faraday rotation are the expansion of 
the azimuthal patterns in Faraday rotation into Fourier series in azimuthal angle \citep{SSK92,Berkhuijsen+97,Fletcher+04,Fletcher+11}
and wavelet analysis \citep{Frick+00,Patrikeev+06}.

In this paper we compare predictions of galactic dynamo theory with the pitch angles of the large-scale magnetic
fields in those galaxies where observational data have been interpreted appropriately.
The pitch angle of the mean magnetic field is defined as $p=\arctan(\mbr/\mbp)$, with $-90^\circ<p\le90^\circ$,
where $(r,\phi,z)$ are cylindrical polar coordinates
with the $z$-axis parallel to the galactic angular velocity $\bfOmega$,
and $\meanv{B}$ is the regular magnetic field.
This is the acute angle between the mean magnetic field and the tangent to the circumference in the
galactic plane, and its negative values signify a trailing spiral. There are several advantages
in using the magnetic pitch angle to test the galactic dynamo theory \citep{Baryshnikova+87,Krasheninnikova+89}.
Firstly, $p$ is readily predicted by the standard non-linear mean-field dynamo theory,
with the need to fix the values of only a small number of parameters
\defcitealias{Chamandy+14b}{CSSS}\defcitealias{Chamandy+Taylor15}{CT}
(\citealt{Sur+07b},\citealt{Chamandy+14b} (hereafter \citetalias{Chamandy+14b}),
\citealt{Chamandy+Taylor15} (hereafter \citetalias{Chamandy+Taylor15})).
Secondly, the closely related position angle of the polarized synchrotron emission is directly observable,
though it must be corrected for Faraday rotation.
In contrast, the \textit{strength} of the mean magnetic field 
depends on additional parameters involving less certain physics.
Its observational determination relies on the questionable assumption of an energy or pressure equipartition
between cosmic rays and magnetic fields \citep{BK05,Stepanov+14} when deduced from the synchrotron intensity,
or the implicit assumption of the statistical independence of the magnetic, 
thermal-electron and cosmic-ray density fluctuations when obtained from Faraday rotation \citep{Beck+03}.
As a result, the regular magnetic field magnitude $\mean{B}$ determined from observational data 
is arguably less reliable than its inferred pitch angle.
Moreover, $\mean{B}$ is sensitive to details of the dynamo non-linearity
and the extent to which the dynamo is supercritical,
to which $p$ is less sensitive.

\citet{Fletcher10} (see also \citealt{Klein+Fletcher15}),
briefly assessed the ability of dynamo theory to explain some features of galactic magnetic fields, but
they restricted the comparison to the kinematic, or linear regime of dynamo action where magnetic field
still grows exponentially. Nearby galaxies definitely have the dynamo action saturated since
the energy density of the large-scale magnetic field is comparable to the turbulent energy density, so 
nonlinear, steady-state solutions of the dynamo equations are more relevant in such comparisons.
As pointed out by \citet{Elstner05} and studied in detail by \citetalias{Chamandy+Taylor15},
magnetic pitch angles are predicted to be smaller in magnitude in the saturated state
than in the kinematic regime.
\citetalias{Vaneck+15} expanded the database compiled by \citet{Fletcher10} and performed a statistical analysis of the data,
including an assessment of the level of agreement between the data and solutions of a simple analytical non-linear dynamo model.
The magnetic pitch angles predicted by the dynamo model used by \citetalias{Vaneck+15} 
had magnitudes much too small to explain observations,
confirming concerns voiced previously \citep{Elstner05}.
\citetalias{Chamandy+Taylor15} explored the parameter space of the dynamo models more extensively
to find magnetic pitch angles comparable to those observed can be obtained in standard dynamo models 
but not with the canonical, solar-neighborhood parameter values used by \citetalias{Vaneck+15}.
Our goal here is to develop further such a comparison using 
essentially the same dynamo model as \citetalias{Chamandy+Taylor15} (with a few refinements).
Given the deliberate simplicity of the galactic dynamo models used, the incompleteness of the
galactic dynamo theory in general, and imperfections in the observational data, it is unrealistic to 
expect an agreement between the theory and observations at the level of rigorous statistical tests,
and such comparisons would necessarily remain qualitative at least until more extensive, homogeneous and reliable
observational data become available.

The paper is organized as follows.
In Section~\ref{sec:data}, we briefly discuss the data set.
We then present dynamo solutions in Section~\ref{sec:model}
illustrating both numerical and analytical results 
in various regions of the parameter space.
Sections~\ref{sec:data} and \ref{sec:model} present only a brief discussion,
and the reader is referred to \citetalias{Vaneck+15} and \citetalias{Chamandy+Taylor15} for more detail.
A discussion of the galactic model used as input to the model can be found in Section~\ref{sec:input}.
Section~\ref{sec:output} presents the main results,
comparing the magnetic pitch angles obtained in the dynamo theory and observations,
while Section~\ref{sec:variations} explores effects of varying model parameters.
We discuss the implications of our results 
and the significance of the remaining discrepancies between the theory and the data in Section~\ref{sec:discussion}.
We formulate our conclusions in Section~\ref{sec:conclusions}.

\begin{table*}
  \begin{center}
    \caption{\label{tab:B_data}Large-scale magnetic fields observed in the sample galaxies:
             the galactocentric radius, magnetic pitch angle $p$, and strength of the axisymmetric component 
             of the mean magnetic field $\mean{B}$ are given in Columns $2$--$4$ \citepalias[][and references therein]{Vaneck+15}.
             Columns $5$--$7$ refer to saturated solutions of our fiducial numerical model (`fid') 
             with $\tau=14\Myr$ and exponentially flared disks 
             (see Figures~\ref{fig:p}, \ref{fig:B}, and \ref{fig:growth_time} 
             for a graphical representation of $p$, $B$, and $T_3$, respectively.)
             The quantity $T_3$ is the $10^3$-folding time of the mean magnetic field in the kinematic (linear) regime.
             Column $8$ shows results for $p$ for the model for which $\tau$ is allowed to vary from galaxy to galaxy 
             (`$\tau$ var.'), discussed in Section~\ref{sec:variable_tau}:
             $\tau=10\Myr$ for M31, $14\Myr$ for M33, $20\Myr$ for M51, $30\Myr$ for M81, $16\Myr$ for NGC~253, 
             $19\Myr$ for NGC~1097, $12\Myr$ for NGC~1365, $10\Myr$ for NGC~1566, $14\Myr$ for NGC~6946, and $11\Myr$ for IC~342.
             Column $9$ refers to the model with $h=\const=0.4\kpc$ (`$h$ const.'), 
             which is discussed in Section~\ref{sec:constant_h}.
             Finally, columns $10$ and $11$ respectively show results for the models which have no outflow (`$U\f=0$'),
             and for which the outflow of the fiducial model was multiplied by $10$ (`$U\f=10\times$'),
             both of which are discussed in Section~\ref{sec:outflows}.
             These differ from the fiducial model only for galaxies for which data were 
             available to estimate the outflow velocity: M31, M33, M51, NGC~253, and NGC~6946.
            }
    \begin{tabular}{@{}ccccccccccc@{}}
      \hline
      \hline
      (1)				&(2)						&(3)		&(4)				&(5)		&(6)			&(7)		&(8)					&(9)				&(10)				&(11)\\
       Galaxy  &Radial range   &$p$   &$\mean{B}$  &$p$   &$\mean{B}$ &$T_3$  &$p$          &$p$          &$p$       &$p$              \\
               &               &(obs.)&(obs.)      &(fid.)&(fid.)     &(fid.) &($\tau$ var.)&($h$ const.) &($U\f=0$) &($U\f=10\times$) \\
               &$[\!\kpc]$     &$[^\circ]$ &$[\!\mkG]$ &$[^\circ]$ &$[\!\mkG]$ &$[\!\Gyr]$ &$[^\circ]$ &$[^\circ]$ &$[^\circ]$ &$[^\circ]$  \\
      \hline                                                                                                                            
      M31      &$6$--$8$    &$-13\pm4$ &$4.8$ &$-28.7$ &$0.9$ &$1.1$ &$-19.9$  &$-17.4$   &$-27.6$   &$-31.5$     \\
               &$8$--$10$   &$-19\pm3$ &$5.6$ &$-18.6$ &$1.1$ &$1.2$ &$-13.2$  &$-16.0$   &$-17.8$   &$-20.3$     \\
               &$10$--$12$  &$-11\pm3$ &$4.7$ &$-15.3$ &$1.1$ &$1.7$ &$-10.9$  &$-18.4$   &$-14.7$   &$-16.5$     \\
               &$12$--$14$  &$-8\pm3$  &$4.9$ &$-13.7$ &$0.8$ &$2.5$ &$-9.8$   &$-23.0$   &$-13.2$   &$-14.6$     \\[7pt]
      M33      &$1$--$3$    &$-51\pm2$ &$0.7$ &$-56.0$ &$1.2$ &$5.4$ &$-56.0$  &$-19.6$   &$-55.4$   &$-59.1$     \\
               &$3$--$5$    &$-41\pm2$ &$0.3$ &$-26.0$ &$1.1$ &$3.2$ &$-26.0$  &$-24.0$   &$-25.5$   &$-27.2$     \\[7pt]
      M51      &$2.4$--$3.6$  &$-20\pm1$ &$1.2$ &$-12.5$ &$6.4$ &$0.19$ &$-18.2$ &$-4.8$   &$-11.8$   &$-13.2$   \\
               &$3.6$--$4.8$  &$-24\pm4$ &$1.5$ &$-15.3$ &$4.8$ &$0.29$ &$-22.4$ &$-7.9$   &$-14.5$   &$-16.2$   \\
               &$4.8$--$6.0$  &$-22\pm4$ &$2.5$ &$-17.6$ &$3.8$ &$0.54$ &$-25.9$ &$-12.0$  &$-16.7$   &$-18.6$   \\
               &$6.0$--$7.2$  &$-18\pm1$ &$2.5$ &$-13.5$ &$2.8$ &$0.71$ &$-19.6$ &$-12.3$  &$-12.8$   &$-14.4$   \\[7pt]
      M81      &$6$--$9$      &$-21\pm7$ &--    &$-7.8$  &$1.7$ &$0.96$ &$-17.0$   &$-11.9$   &$-7.8$   &$-7.8$    \\
               &$9$--$12$     &$-26\pm6$ &--    &$-4.6$  &$0.8$ &$2.1$  &$-9.9$   &$-16.1$   &$-4.6$   &$-4.6$    \\[7pt]
      NGC~253  &$1.4$--$6.7$  &$-25\pm5$ &$4.4$ &$-21.8$ &$2.8$ &$0.40$ &$-25.3$   &$-10.0$   &$-20.6$   &$-24.7$  \\[7pt]
      NGC~1097 &$1.25$--$2.50$  &$-34\pm2$ &$1.4$ &$-8.1$ &$7.7$ &$0.12$ &$-11.2$   &$-2.1$   &$-8.1$   &$-8.1$    \\
               &$2.50$--$3.75$  &$-36\pm4$ &$1.1$ &$-11.4$ &$5.9$ &$0.16$ &$-15.7$  &$-3.4$   &$-11.4$  &$-11.4$ \\
               &$3.75$--$5.00$  &$-23\pm2$ &$1.9$ &$-22.0$ &$3.9$ &$0.27$ &$-31.5$  &$-7.4$   &$-22.0$  &$-22.0$ \\[7pt]
      NGC~1365 &$2.625$--$4.375$   &$-34\pm2$ &$0.8$ &$-46.9$ &$2.4$ &$0.93$ &$-37.3$ &$-11.0$ &$-46.9$ &$-46.9$     \\
               &$4.375$--$6.125$   &$-17\pm1$ &$0.8$ &$-28.9$ &$3.6$ &$0.40$ &$-24.2$ &$-9.2$  &$-28.9$ &$-28.9$     \\
               &$6.125$--$7.875$   &$-31\pm1$ &$0.7$ &$-29.6$ &$3.1$ &$0.78$ &$-24.8$ &$-11.3$ &$-29.6$ &$-29.6$     \\
               &$7.875$--$9.625$   &$-22\pm1$ &$1.2$ &$-29.9$ &$2.5$ &$1.5$  &$-25.0$ &$-13.7$ &$-29.9$ &$-29.9$     \\
               &$9.625$--$11.375$  &$-37\substack{+6\\-1}$ &$1.1$ &$-28.6$  &$1.9$ &$2.7$ &$-24.0$ &$-15.9$ &$-28.6$ &$-28.6$     \\
               &$11.375$--$13.125$ &$-29\pm11$ &$0.7$ &$-28.0$ &$1.1$ &$5.8$ &$-23.5$ &$-18.9$ &$-28.0$ &$-28.0$  \\
               &$13.125$--$14.875$ &$-33\pm6$ &$0.4$ &$-27.8$ &$0.5$ &$25.$ &$-23.4$ &$-22.8$ &$-27.8$ &$-27.8$   \\[7pt]
      NGC~1566 &$2.7$--$8.0$ &$-20\pm5$ &--  &$-28.2$ &$3.4$ &$0.88$ &$-19.4$ &$-12.7$ &$-28.2$ &$-28.2$     \\[7pt]
      NGC~6946 &$0$--$6$     &$-27\pm2$ &--  &$-29.4$ &$3.1$ &$0.55$ &$-29.4$ &$-11.9$ &$-28.2$ &$-31.0$     \\
               &$6$--$12$    &$-21\pm2$ &--  &$-12.6$ &$0.8$ &$2.9$  &$-12.6$ &$-24.5$ &$-12.0$ &$-14.7$     \\
               &$12$--$14$   &$-10\pm6$ &--  &$-7.1$  &$0.4$ &$6.2$  &$-7.1$  &$-40.0$ &$-6.5$  &$-9.8$     \\[7pt]
      IC~342   &$5$--$9$     &$-21\pm2$ &--  &$-26.4$ &$1.8$ &$1.8$ &$-20.2$  &$-16.1$ &$-26.4$ &$-26.4$     \\
               &$9$--$13$    &$-18\pm2$ &--  &$-23.6$ &$0.6$ &$9.7$ &$-18.2$  &$-29.9$ &$-23.6$ &$-23.6$     \\
      \hline\hline
    \end{tabular}
  \end{center}
\end{table*}
\begin{table*}
  \begin{center}
    \caption{\label{tab:input_data}Observational data used as input into the model to calculate the configuration of the mean magnetic field.
             Data from columns 2--6 are taken from \citetalias{Vaneck+15}.
             Column~3 gives the magnitude of the angular velocity $\Omega$ averaged over the 
             radial range shown in column~2.
             Column~4 gives the shear parameter $q=-S/\Omega$ averaged over the radial bin,
             where the radial shear $S$ is taken from \citetalias{Vaneck+15}.
             Columns~5 and 6 refer to the HI and star formation rate surface densities, respectively
             (if the data were not available, the entry reads `--').
             The corrected value of $d_{25}=2r_{25}$ is taken as the mean of LEDA (column~7) and NED (column~8) database entries.
             Those values denoted with `$*$' in the outermost radial bin of NGC~6946 are extrapolations 
             assuming exponential profiles for $\Sigma_I$ and $\Sigma_\mathrm{*}$.
            }
    \begin{tabular}{@{}cccccccc@{}}
      \hline
      \hline
      (1)				&(2)								&(3)					&(4)			&(5)							&(6)													&(7)						&(8)\\
       Galaxy  &Radial range       &$\Omega$      &$q$     &$\Sigma_I$        &$\Sigma_\mathrm{*}$         &$\log(2r_{25})$ &$\log(2r_{25})$ \\
               &                   &              &        &                  &                            &LEDA            &NED             \\
               &$[\!\kpc]$         &$[\!\kmskpc]$ &        &$M_\odot\pc^{-2}$ &$M_\odot\pc^{-2}\Gyr^{-1}$  &$[0.1']$        &$[0.1']$        \\
      \hline                                                                                                                              
      M31      &$6$--$8$           &$38.4$        &$0.75$  &$1.47$            &$0.443$                     &$3.28$          &$3.31$          \\
               &$8$--$10$          &$31.1$        &$0.99$  &$2.17$            &$0.621$                     &                &                \\
               &$10$--$12$         &$25.1$        &$1.07$  &$3.64$            &$0.794$                     &                &                \\
               &$12$--$14$         &$21.1$        &$1.02$  &$4.05$            &$0.227$                     &                &                \\[7pt]
      M33      &$1$--$3$           &$40.7$        &$0.62$  &$11.3$            &$9.64$                      &$2.80$          &$2.87$          \\
               &$3$--$5$           &$24.9$        &$0.84$  &$9.43$            &$3.99$                      &                &                \\[7pt]
      M51      &$2.4$--$3.6$       &$86.5$        &$1.13$  &$5.20$            &$20.7$                      &$2.15$          &$2.05$          \\
               &$3.6$--$4.8$       &$58.1$        &$1.05$  &$5.97$            &$13.5$                      &                &                \\
               &$4.8$--$6.0$       &$46.7$        &$0.87$  &$8.98$            &$18.0$                      &                &                \\
               &$6.0$--$7.2$       &$38.1$        &$1.05$  &$6.63$            &$7.92$                      &                &                \\[7pt]
      M81      &$6$--$9$           &$31.7$        &$1.24$  &$3.33$            &--                          &$2.36$          &$2.44$          \\
               &$9$--$12$          &$19.8$        &$1.49$  &$2.32$            &--                          &                &                \\[7pt]
      NGC~253  &$1.4$--$6.7$       &$50.9$        &$0.97$  &$3.28$            &$35.1$                      &$2.40$          &$2.45$          \\[7pt]
      NGC~1097 &$1.25$--$2.50$     &$182$         &$1.20$  &$3.00$            &--                          &$2.03$          &$1.98$          \\
               &$2.50$--$3.75$     &$94.5$        &$1.43$  &$2.95$            &--                          &                &                \\
               &$3.75$--$5.00$     &$62.1$        &$1.00$  &$3.13$            &--                          &                &                \\[7pt]
      NGC~1365 &$2.625$--$4.375$   &$71.3$        &$0.58$  &$3.39$            &--                          &$2.08$          &$2.05$          \\
               &$4.375$--$6.125$   &$52.4$        &$0.95$  &$4.31$            &--                          &                &                \\
               &$6.125$--$7.875$   &$39.3$        &$1.05$  &$6.89$            &--                          &                &                \\
               &$7.875$--$9.625$   &$30.9$        &$1.11$  &$9.50$            &--                          &                &                \\
               &$9.625$--$11.375$  &$25.1$        &$1.19$  &$10.8$            &--                          &                &                \\
               &$11.375$--$13.125$ &$20.9$        &$1.22$  &$9.58$            &--                          &                &                \\
               &$13.125$--$14.875$ &$17.7$        &$1.21$  &$8.26$            &--                          &                &                \\[7pt]
      NGC~1566 &$2.7$--$8.0$       &$38.2$        &$0.96$  &$9.69$            &--                          &$1.86$          &$1.92$          \\[7pt]
      NGC~6946 &$0$--$6$           &$48.1$        &$0.86$  &$6.30$            &$20.2$                      &$2.19$          &$2.22$          \\
               &$6$--$12$          &$19.6$        &$1.05$  &$4.08$            &$2.50$                      &                &                \\
               &$12$--$14$         &$13.4$        &$1.02$  &$3.05^*$          &$0.620^*$                   &                &                \\[7pt]
      IC~342   &$5$--$9$           &$28.1$        &$1.05$  &$6.41$            &--                          &$2.62$          &$2.65$          \\
               &$9$--$13$          &$17.7$        &$0.95$  &$6.53$            &--                          &                &                \\
      \hline\hline
    \end{tabular}
  \end{center}
\end{table*}

\section{Data}
\label{sec:data}
Our sample contains ten nearby galaxies selected because observations of their synchrotron emission and, especially,
Faraday rotation, have been interpreted with sufficient detail and reliability in terms of the pitch angle of the
large-scale magnetic field as to admit comparison with theory. In particular, the azimuthal variation of the synchrotron
polarization angle in most of them has been interpreted in terms of the magnitudes of the Fourier components
of the product $n_\mathrm{e}\vec{B}h$, where $n_\mathrm{e}$ is the thermal electron density and $h$ is the scale height
of the ionized layer. This allows a reliable estimate of the pitch angle of the axisymmetric part of the large-scale magnetic
field (the azimuthal wave number $m=0$) whose parameters can be predicted by the dynamo theory. The mode $m=1$ corresponds
to a bisymmetric structure, most often associated with an overall asymmetry of the pattern rather than
dynamo action, whereas the mode $m=2$ is usually attributed to a perturbation induced by a two-armed spiral pattern. 
In all cases where magnetic field structure is known for both the disk and halo, we use the disk data.

Table~\ref{tab:B_data} presents parameters of the magnetic fields in the sample galaxies, both as observed and as obtained
from various versions of the dynamo models discussed in Section~\ref{sec:model}. 
Table~\ref{tab:input_data} contains the galactic parameters used as an input to the dynamo models. 
We mostly use the data compiled by \citetalias{Vaneck+15} but with a few notable exceptions.
A typographical error in the magnetic pitch angle of IC~342 has been corrected:
in the outermost annulus, $p=-18^\circ\pm2^\circ$ \citep[c.f.][]{Grave+Beck88}.
We exclude the galaxies M94 and NGC~4414 from the sample
since the radial range over which the data is averaged 
is not stated in the original work \citepalias{Vaneck+15}.
NGC~253 is included, however, with the radial range $1.4$--$6.7\kpc$ \citep{Heesen+09b}.
For M33, we have selected the observational model from \citet{Tabatabaei+08}
that contains $m=0$ and $m=2$ azimuthal modes but no vertical field component $\mbz$,
as this type of magnetic field is more likely to be maintained in a thin gas layer
(this choice is different from that of \citetalias{Vaneck+15}).
To determine the \ion{H}{1} and star formation rate surface densities in 
the outer annulus of NGC~6946, exponentials were fitted to the data from the two innermost annuli.

We divide the sample into three groups based on the galaxy morphology and magnetic field fitting technique.
Group~I consists of M31, M33, and M51, galaxies without a prominent bar, for which azimuthal 
Fourier components ($m=0$, $m=1$, and $m=2$) have been fitted to the multi-frequency data on the
synchrotron polarization angle in the galactic disk.
Following \citetalias{Vaneck+15}, we compare our dynamo model with the axisymmetric ($m=0$) component.
Group~II consists of M81, NGC~253, NGC~1566, NGC~6946, and IC~342.
M81 has no bar whereas the others are SAB galaxies but with a small bar.
In each case, the bar does not extend to the midpoint of the inner radial range considered.
The bar radii are about $2.2\kpc$ for NGC~253 \citep{Sorai+00},
$1.7\kpc$ for NGC~1566 \citep{Combes+14}, $1.8\kpc$ for NGC~6946 \citep[][and references therein]{Fathi+07},
and $1.5\kpc$ for IC~342 \citep{Crosthwaite+00}.
For M51 and NGC~253, disk and halo magnetic field components were fitted separately \citep{Fletcher+11,Heesen+09b},
while for M31, such a separation was considered but found to be unnecessary \citep{Fletcher+04}.
Finally, Group~III consists of two barred galaxies,
NGC~1097 and NGC~1365, with bars as large as about $10\kpc$ and $11\kpc$, respectively, thus
extending into most (NGC~1365) or the whole (NGC~1097) of the range of galactocentric distances
where the pitch angle of the mean magnetic field is known \citep{Beck+05}.
The structure of the mean magnetic field in and near the bar
is affected by the associated strongly nonaxisymmetric gas flow \citep{Moss+01,Moss+07},
so we do not expect our dynamo model, designed for axially symmetric galaxies, to be very successful 
at explaining such fields.

\section{Dynamo model}
\label{sec:model}
The dynamo model is essentially that of \citetalias{Chamandy+Taylor15}
and \citetalias{Chamandy+14b},
with a few refinements, and we refer the reader to those papers for more information, but
here we introduce it only briefly.
We solve three coupled dynamo equations in the slab approximation,
where radial derivatives are neglected.
This approximation has been shown to produce accurate results
when compared to higher dimensional solutions \citep{Chamandy16}.
The radial and azimuthal components of the mean-field induction equation
and the dynamical quenching equation, which stems from magnetic helicity balance \citep{Shukurov+06},
are then given by
\begin{align}
  \label{dBrdt}
  \frac{\del\mbr}{\del t}=& -\frac{\del}{\del z}\left(\alpha\mbp\right) 
                            +\eta\frac{\del^2\mbr}{\del z^2} -\frac{\del}{\del z}\left(\muz\mbr\right),\\
  \label{dBpdt}
  \frac{\del\mbp}{\del t}=& -q\Omega\mbr 
                            +\frac{\del}{\del z}\left(\alpha\mbr\right) 
                            +\eta\frac{\del^2\mbp}{\del z^2} -\frac{\del}{\del z}\left(\muz\mbp\right),\\
  \label{dalp_mdt}
  \frac{\del\alpha\magn}{\del t}=& -\frac{2\eta}{l^2 B\eq^2}
                                     \left[ \alpha\left(\mbr^2+\mbp^2\right)
                                     -\eta\left(\frac{\del\mbr}{\del z}\mbp 
                                     				\phantom{\frac{\del\mbr}{\del z}}
                                     				\right.\right.\nonumber\\
                                   & \left.\left.-\frac{\del\mbp}{\del z}\mbr \right)\right]
                                    -\frac{\del}{\del z}\left(\muz\alpha\magn\right) +\kappa\frac{\del^2\alpha\magn}{\del z^2}.
\end{align}
with $l$ the turbulent correlation length, 
$\alpha$ equal to the sum of kinetic and magnetic contributions,
\[
  \alpha=\alpha\kin+\alpha\magn,
\]
and the field strength corresponding to energy equipartition with turbulent kinetic energy given by
\[
  B\eq= u\sqrt{4\pi\rho},
\]
with $\rho$ the density of ionized gas and $u$ is the rms turbulent velocity.
The turbulent diffusivities $\eta$ and $\kappa$ are assumed to be independent of $z$.
We adopt the standard estimate
\[
  \eta= \frac{1}{3}\tau u^2,
\]
with both $u$ and the correlation time of interstellar turbulence $\tau$ 
assumed constant.
The mean vertical velocity and kinetic $\alpha$ effect are given by
\[
  \muz= U\f\frac{z}{h}
\]
and
\[
  \alpha\kin= \alpha\f\sin\left(\frac{\pi z}{h}\right),
\]
respectively,
with $U\f$ and $\alpha\f$ both parameters with dimensions of velocity.
Note that we have neglected $\mbz\del\mup/\del z$ in equation~\eqref{dBpdt}, 
and assumed $\mbz^2\ll\mbr^2+\mbp^2$ in equation~\eqref{dalp_mdt},
which are appropriate for a thin disk.
We have also assumed $\mur=0$.
Vacuum boundary conditions, $\mbr=\mbp=0$ at $z=\pm h$, are used,
along with $\del^2\alpha\magn/\del z^2=0$ at $z=\pm h$.
The boundary condition on $\alpha\magn$ leaves unconstrained the helicity flux through the disk surface.

\subsection{Dimensionless parameters}
\label{sec:parameterization}
The following dimensionless parameters 
(all are or can be functions of the galactocentric distance) 
can be used to specify the dynamo model \citepalias{Chamandy+Taylor15}:
\begin{equation}
  \label{dimensionless}
  q=-\frac{d\ln\Omega}{d\ln r}, \quad
  H= \frac{h}{\tau u}, \quad
  \Coriolis\equiv\Omega\tau, \quad
  V= \frac{U\f}{u},
\end{equation}
where $\Omega$ is the galactic angular velocity, $r$ is the galactocentric distance,
$h$ is the scale height of the magnetized gas layer, and $U\f$ is the characteristic
speed of the galactic outflow (wind or fountain). The velocity shear due to the galactic differential 
rotation is quantified by $q$,
$H$ is the disk half-thickness in the unit $\tau u$ comparable to the turbulent correlation length,
$\Coriolis$ is the Coriolis number, a measure of the influence of rotation on the interstellar turbulence,
and $V$ is the outflow speed at the disk surface $U\f$ in the units of the turbulent speed $u$.

\begin{figure}
  \begin{centering}
  \includegraphics[width=0.6\columnwidth,clip=true,trim=0 0  0 0]{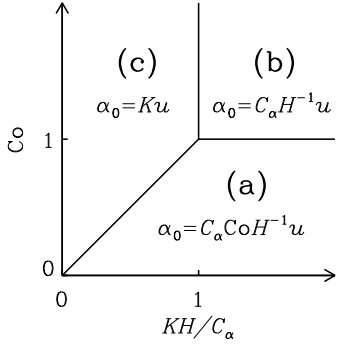}	
  \caption{A diagram of the parameter space showing where each of the cases of constraint~\eqref{constraint}
           are applicable.          \label{fig:pspace}
          }            
  \end{centering}
\end{figure}

The most significant refinement to the dynamo model of \citetalias{Chamandy+Taylor15} 
is that we adopt the amplitude of the (kinetic part of the)
$\alpha$ coefficient as \citep[][p.~163]{Ruzmaikin+88}
\begin{subnumcases}{\label{alpha0}  \alpha\f=\frac{C_\alpha\tau u^2}{h}\times}
    \tau\Omega, 	&if  $\Coriolis\le1,$\label{a0a}\\
    1, 						&if  $\Coriolis>1,$\label{a0b}
\end{subnumcases}
where $C_\alpha$ is a constant of order unity. In terms of the dimensionless parameters,
\[
  \alpha\f=
  \frac{C_\alpha u}{H}\times
  \begin{dcases}
    \Coriolis, 	&\mbox{if }  \Coriolis\le1,\\
    1, 					&\mbox{if } \Coriolis>1.
  \end{dcases}
\]
In both cases, $\alpha\f\le Ku$, with a certain constant $K$ of order unity, because $\alpha\f$ 
cannot exceed some fraction of the turbulent speed. 
We vary parameters $K$ and $C_\alpha$, that allow for uncertainties of the theory, 
to assess the sensitivity of the results to this variation. These estimates of $\alpha\f$ are illustrated in
Figure~\ref{fig:pspace}, where the $(\Co,H)$ parameter plane is divided into three regions, labeled (a), (b) and (c),
where various forms of $\alpha\f$ apply, with (a) and (b) corresponding to Equations~\eqref{a0a} and \eqref{a0b}, 
respectively, and (c) corresponding to $\alpha\f=Ku$.

Alternatively, but equivalently, one could use the turbulent Reynolds numbers $R_\alpha$, $R_\Omega$ and $R_U$, 
in place of $H$, $\Coriolis$, and $V$, as is more conventional in dynamo theory. They are related as
\[
  R_\alpha=
  \begin{dcases}
    3C_\alpha\Coriolis, 	&\mbox{if }  \Coriolis\le1,\\
    3C_\alpha, 			&\mbox{if } \Coriolis>1,
  \end{dcases}
\]
with the maximum value $R_\alpha=3KH$ to prevent $\alpha\f$ from exceeding $Ku$.
In addition, we have
\[
  R_\Omega= -3q\Coriolis H^2
\qquad\text{and}\qquad
  R_U= 3HV.
\]
 
The dynamo action saturates, and the system settles to a steady state, when the energy density
of the mean magnetic field has grown to become comparable (but not equal) to the turbulent
energy density, i.e., when the magnetic field strength becomes comparable to $B\eq$.
In numerical solutions of the dynamo equations, which resolve the $z$-dependence of the galactic disk and
magnetic field, we adopt an exponential profile for $B\eq$ as a function of $z$,
with scale height $h$.
The pitch angle determined from numerical solutions
is calculated by taking a weighted vertical average with respect to $B\eq^2\mean{B}^2$,
as the intensity of polarized synchrotron emission is proportional 
to the cosmic ray electron density,
which is expected to scale with $B\eq^2$ 
(where $B\eq$ is an approximation of the total field strength comprising both the mean and random parts)
at the kiloparsec scales \citep{Stepanov+14}.
As $B\eq$ is largest at the midplane, where $|p|$ is also typically largest, 
this shifts the weighted average to slightly larger values of $|p|$.
Likewise, averages of mean field strength $B^2$ across the disk are weighted by $B\eq^2$
to allow for the fact that they are often obtained from the synchrotron intensity.

Below we use both analytical and numerical solutions for the mean magnetic field,
both obtained in the local (slab) approximation.
Given the accuracy of the analytical solutions that we demonstrate, they 
enable deeper insight into the theory and results and facilitate further analysis.

\subsection{Analytical dynamo solutions}
\label{sec:analytical}
Simple and yet accurate analytical solutions of the galactic mean-field dynamo equations 
can be obtained  by approximating the $z$-derivatives of magnetic field as divisions by $h$ with appropriate 
numerical coefficients (the no-$z$ approximation). Here we present the results
obtained from this model referring the reader to \citepalias{Chamandy+14b} for details.
To obtain the steady-state solution for the saturated state, one assumes that 
\citetalias{Chamandy+14b,Chamandy+Taylor15},
\begin{equation}
  \label{alpha_crit}
  \alpha=\alpha\crit= \left(\frac{\pi}{2}\right)^5\left(1+\frac{6HV}{\pi^2}\right)^2\frac{u}{9q\Coriolis H^3},
\end{equation}
and then the pitch angle is given by
\begin{equation}
  \label{psat}
  \tan p\sat= -\frac{\pi^2 +6HV}{12q\Coriolis H^2}.
\end{equation}
Furthermore, the admissible region in the parameters space is constrained by the requirement that
the dynamo action can maintain a large-scale magnetic field, i.e., $\alpha\f\ge\alpha\crit$,
where $\alpha\f$ is given in Eq.~\eqref{alpha0}. Straightforward algebra then leads to
the admissible values of the Coriolis number,
  \begin{subnumcases}{\label{constraint}\Coriolis\ge \frac{\psi}{H}\times}
    \frac{1}{C_\alpha^{1/2}},\!\!\!\!\!\!  &if $\displaystyle\min\left(1,\frac{KH}{C_\alpha}\right)\ge\Coriolis,$\label{CoI}\\
    \frac{\psi}{C_\alpha H},\!\!\!\!\!\!   &if $\displaystyle\min\left(\Coriolis,\frac{KH}{C_\alpha}\right)\ge 1,$\label{CoII}\\
    \frac{\psi}{K H^2},\!\!\!\!\!\!        &if $\displaystyle\min\left(\Coriolis,1\right)\ge\frac{KH}{C_\alpha},$\label{CoIII}
  \end{subnumcases}
where
\begin{equation}
  \label{psi}
  \psi= \frac{1}{12}\sqrt{\frac{\pi}{2q}}\left(\pi^2+6HV\right).
\end{equation}
These cases correspond to the regions of parameter space shown in Figure~\ref{fig:pspace}.
Yet another requirement is that the dynamo amplifies the mean magnetic field fast enough,
i.e., the kinematic growth rate of the mean magnetic field, $\gamma$, is sufficiently large. 
In each of the cases \eqref{CoI}--\eqref{CoIII}, we have, respectively,
\begin{subnumcases}{\label{gamma}\gamma= \sqrt{\frac{2q}{\pi}}\frac{\Omega}{H}\times}
  \sqrt{C_\alpha} -\frac{\psi}{\Coriolis H}, \\
  \sqrt{\frac{C_\alpha}{\Coriolis}} -\frac{\psi}{\Coriolis H}, \\
  \sqrt{\frac{KH}{\Coriolis}} -\frac{\psi}{\Coriolis H}.
\end{subnumcases}
This is a \textit{local} growth rate as it represents the growth rate at a given position 
in the disk unimpeded by the radial and azimuthal magnetic diffusion, so $\gamma$ is a function of 
the galactocentric distance $r$ in an axially symmetric disk.
The diffusive and non-local coupling within the disk 
leads to the establishment of a \textit{global} eigenmode
whose growth rate $\Gamma$ is independent of position and is
only slightly (a few percent in a thin disk) smaller 
than the \textit{largest} value of $\gamma(r)$ in an axisymmetric disk 
\citep[Section VII.6 in][]{Ruzmaikin+88,Moss+98b}.
We require solutions to have $\gamma>0$ as only those solutions produce non-zero mean magnetic field.

Finally, the strength of the mean magnetic field in the saturated dynamo regime is given by
\begin{equation}
  \label{Bsat}
  B\sat^2= \frac{B\eq^2\Strouhal^2}{2\xi H^2}\left(\frac{\tan^2p\kin}{\tan^2p\sat}-1\right)(\pi^2R_\kappa +3HV),
\end{equation}
where $\Strouhal\equiv l/(\tau u)$ is the Strouhal number,
$R_\kappa\equiv \kappa/\eta$ with $\kappa$ the turbulent diffusivity of $\alpha\magn$,
and $\xi\equiv 1-3\cos^2p\sat/(4\sqrt{2})$ (with $\xi\approx0.5$ as $p\sat\rightarrow0$).
Here $p\kin$ is the pitch angle in the kinematic regime, given by
\begin{subnumcases}{\label{pkin}  \tan^2p\kin= \frac{2}{\pi q H}\times}
      \displaystyle\frac{C_\alpha}{H},  \\
      \displaystyle\frac{C_\alpha}{\Coriolis H},\\
      \displaystyle\frac{K}{\Coriolis},
\end{subnumcases}
in the cases \eqref{CoI}--\eqref{CoIII}, respectively.
Note that, unlike $p\sat$, outflows do not affect the value of $p\kin$ in the analytical solution.
A property of the solution is that $p\kin/p\sat\ge1$.
It is also worth noting that \citepalias[][e.g.]{Chamandy+14b,Chamandy+Taylor15},
\begin{equation*}
  \frac{\tan^2p\kin}{\tan^2p\sat}=\frac{\alpha\f}{\alpha\crit}=\frac{D\f}{D\crit},
\end{equation*}
where $D\f$ and $D\crit$ are the values of the dynamo number $D=-\alpha q\Omega h^3/\eta^2$ 
in the kinematic and saturated regimes, respectively.

\subsection{Numerical dynamo solutions}
\label{sec:numerical}
Numerical solutions are obtained for equations \eqref{dBrdt}--\eqref{dalp_mdt} in $z$ and $t$.
We used a uniform mesh in $z$ with a 6th order finite-difference and 3rd order time-stepping routines 
employed in the Pencil Code \citep{Brandenburg03}.
Solutions are averaged across the disk to obtain a prediction that may be tested against observations
as described in Section~\ref{sec:parameterization}.
We obtained both non-linear and kinematic numerical solutions to derive both the saturated magnetic field and 
the local kinematic growth rate, respectively.
Spatial resolutions of 21 or 41 grid points are generally sufficient to cover the parameter space 
specified in Section~\ref{sec:solution_space},
but in subsequent sections we used a resolution of 101 grid points to increase the precision.

\begin{figure*}                     
  \begin{tabular}{c c c c} 
    \includegraphics[width=0.23\textwidth,clip=true,trim= 20 15 15 10]{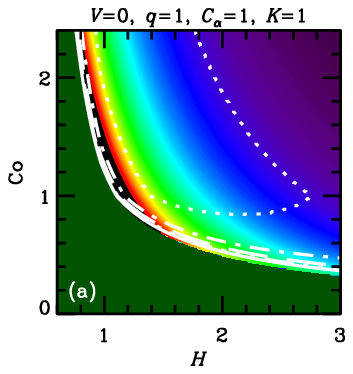}&	
    \includegraphics[width=0.23\textwidth,clip=true,trim= 20 15 15 10]{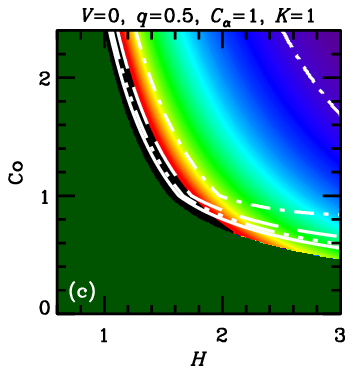}&
    \includegraphics[width=0.23\textwidth,clip=true,trim= 20 15 15 10]{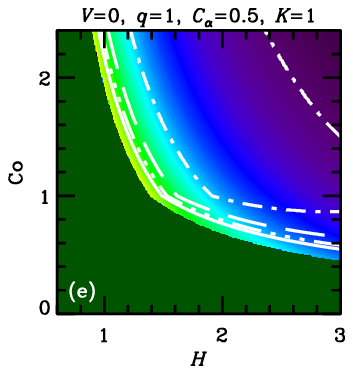}&
    \includegraphics[width=0.23\textwidth,clip=true,trim= 20 15 15 10]{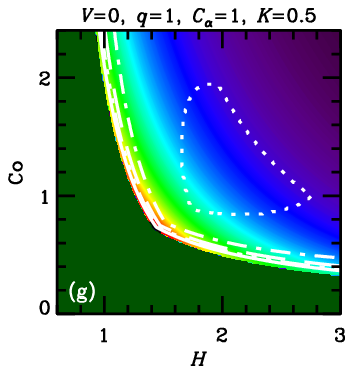}\\
    \includegraphics[width=0.23\textwidth,clip=true,trim= 20 15 15 10]{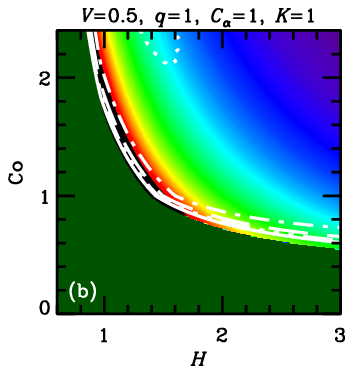}&
    \includegraphics[width=0.23\textwidth,clip=true,trim= 20 15 15 10]{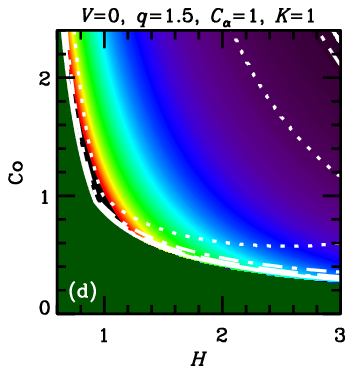}& 
    \includegraphics[width=0.23\textwidth,clip=true,trim= 20 15 15 10]{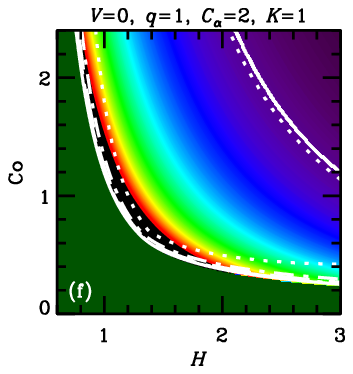}&   
    \includegraphics[width=0.23\textwidth,clip=true,trim= 20 15 15 10]{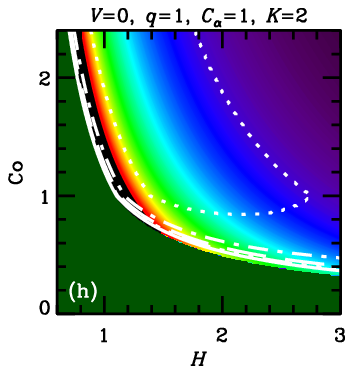}\\
  \end{tabular}
  \begin{center}
    \includegraphics[width=0.7\textwidth,clip=true,trim=-20 20  0 0]{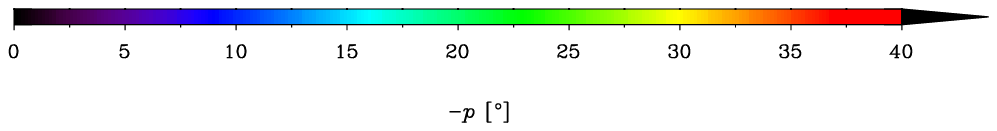}
  \end{center}
  \caption{The pitch angle of the mean magnetic field in the saturated (steady) state (color) for various sets of parameter values 
           given at the top of each panel,
           for the numerical solution, as a function of the dimensionless disk semi-thickness $H$ and Coriolis number $\Coriolis$
           (with each parameter sampled with the step size of $0.01$).
           Contours denote the local $10^3$-folding time in local galactic rotation periods $T_3=6$ (dotted), 
           $12$ (dashed-dotted), $18$ (dashed), $24$ (dash-triple-dotted), and $30$ (solid).
           Panel~(a) shows the solution for fiducial parameter values (no outflow, locally flat rotation curve),
           while panel~(b) includes a strong outflow.
           Each column shows how the solution changes when the value of a single parameter is varied from the fiducial case.
           In the top right corners of panels (d) and (f), the solid contour demarcates the region
           where solutions are oscillatory in the kinematic regime but become stationary upon saturation.
           Dark green denotes the region of parameter space with decaying solutions in the kinematic regime.
           \label{fig:psat_num}
          }            
\end{figure*}
\subsection{Magnetic pitch angle in the dynamo solutions}
\label{sec:solution_space}
In what follows, we discuss the magnetic pitch angle $p$ in the steady state (saturated, nonlinear dynamo 
solutions) unless otherwise specified, and denote it $p$ without a subscript.
The pitch angle of the mean magnetic field obtained from the steady-state solutions of the
dynamo equations \eqref{dBrdt}--\eqref{dalp_mdt} is shown in Figure~\ref{fig:psat_num} for the numerical solution.
For comparison, Figure~\ref{fig:psat_anal} presents $p$ for the analytic solution, equation~\eqref{psat}.
It is evident from these figures that the two solutions are rather similar 
\citepalias[see also][]{Chamandy+14b,Chamandy+Taylor15}.
The pitch angle is shown in the $(H,\Coriolis)$-plane for various combinations of the remaining model
parameters $V$, $q$, $C_\alpha$ and $K$. 
To obtain the numerical solutions, 
we fix $\Omega=40\kmskpc$ and $R_\kappa=1$ (in this section only).
The turbulent outer scale $l$ enters only through the equation for $\alpha\magn$,
and we set $l=\tau u$: that is, we adopt $\Strouhal=1$.
As equations~\eqref{dBrdt}--\eqref{dalp_mdt} can be written 
in terms of the dimensionless parameters \eqref{dimensionless} \citepalias{Chamandy+Taylor15},
the solutions are insensitive to the value of $\Omega$, for a fixed value of $\Coriolis$.
Moreover, in the analytical solution for $p$, which closely approximates numerical solutions,
one simply approximates the dynamo as critical ($\alpha=\alpha\crit$),
without the need to invoke equation~\eqref{dalp_mdt} explicitly.
Thus, we expect solutions for $p$ to be almost independent of $R_\kappa$ and $\Strouhal$;
this is indeed borne out by our numerical solutions.
Also plotted are the contours for the local $10^3$-folding growth time,  
\begin{equation}
  \label{T_3}
  T_3=\ln(10^3)/\gamma.
\end{equation}
In units of the local galactic rotation period $2\pi/\Omega$, 
the dotted contour corresponds to $T_3=6$ and other contours show $T_3=12$, $18$, $24$ and $30$.
Since these growth times refer to the kinematic regime,
when the field is growing exponentially and the non-linearity is unimportant, 
$T_3$ is independent of $R_\kappa$ and $\Strouhal$. 
Regions shown dark green in Figure~\ref{fig:psat_num} 
and Figure~\ref{fig:psat_anal} correspond to decaying solutions.

In each column, a single parameter is varied from the fiducial case ($V=0$, $q=C_\alpha=K=1$) of Panel~(a).
The leftmost columns of Figures~\ref{fig:psat_anal} and \ref{fig:psat_num}
show that, for a given $H$ and $\Coriolis$, a stronger outflow (larger $V$) leads to a larger $|p|$, 
but weakens dynamo action (leading to a larger $T_3$ or a decaying solution) by making
$\alpha\crit$ larger.
Combining equations~\eqref{alpha_crit} and \eqref{psat},
we find $\tan p \propto [\alpha\crit/(hq\Omega)]^{1/2}$,
where the right-hand-side is the ratio of the strengths of $\alpha$ and $\Omega$ effects \citep[c.f.][]{Shukurov07}.
We see then that a larger value of $\alpha\crit$ implies a larger value of $|p|$.

Turning to the next column, we see that a weaker velocity shear (smaller $q$) also causes $|p|$ to increase.
But, again, this suppresses the dynamo action: the dynamo growth rate $\gamma$ decreases with $q$.

In the rightmost two columns, we show the result of varying the parameters $C_\alpha$ and $K$.
The numerical solutions of Figure~\ref{fig:psat_num} confirm that $p$ is almost independent of $C_\alpha$ and $K$,
but that increasing either of these parameters strengthens dynamo action, 
leading to smaller growth times and less stringent constraints on the parameter space.
A solid contour in the top right corner of panel~\ref{fig:psat_num}(f) (also visible in panel (d))
identifies the region where linear eigenmodes are oscillatory, 
with the usual even parity about the midplane, 
but relax to the `expected' stationary solution at saturation
(local growth times are not calculated for those solutions).\footnote{For 
certain other combinations of parameter values, e.g.~$V=0.5$, $q=0.5$, $C_\alpha=2$, and $K=1$,
we find kinematic eigenmodes to be oscillatory with even parity for a certain range of $H$, in this case $H=1.9$--$2.4$.
In most cases, these solutions regularize to the `expected' stationary solutions upon saturation,
but for a small region of parameter space near the boundary where the dynamo becomes critical, 
oscillations can persist into the non-linear regime.
Such cases are probably not realistic, as in radial and azimuthal magnetic diffusion would damp such oscillations,
resulting in stationary global modes.}
Nevertheless, saturated solutions are rather robust to variations of order unity in $C_\alpha$ and $K$.

Comparing Figures~\ref{fig:psat_anal} and \ref{fig:psat_num}, 
we see that analytical solutions slightly underestimate $|p|$ and $\gamma$.
In this respect, the analytical solution is slightly conservative.
The difference is noticeable where the $\alpha^2$ effect is significant,
namely near the boundary of the allowed region of the $(H,\Coriolis)$-plane and for large $V$ or small $q$,
but analytical solutions are, even in the most extreme cases, 
within a factor of two of numerical solutions for the examples presented above
\citepalias[see also][]{Chamandy+14b,Chamandy+Taylor15}.

\begin{figure*}                     
  \includegraphics[width=0.32\textwidth]{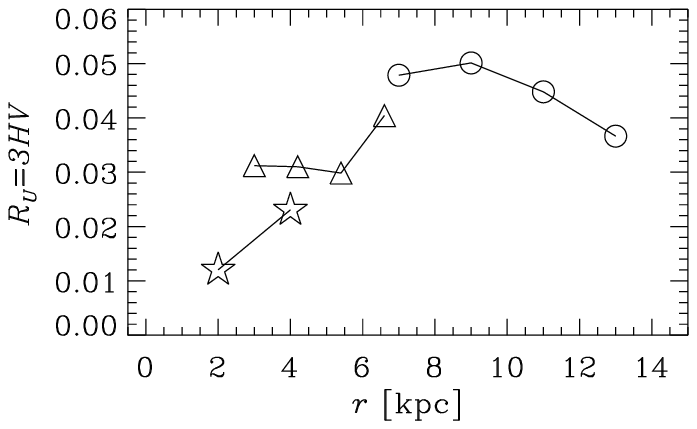}
  \includegraphics[width=0.32\textwidth]{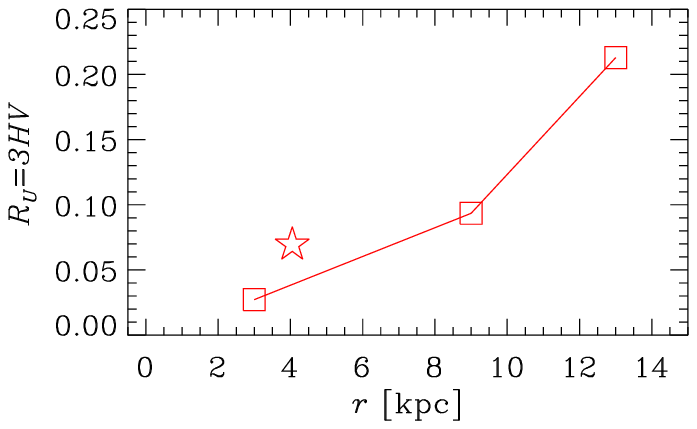}
  \\
  \includegraphics[width=0.32\textwidth]{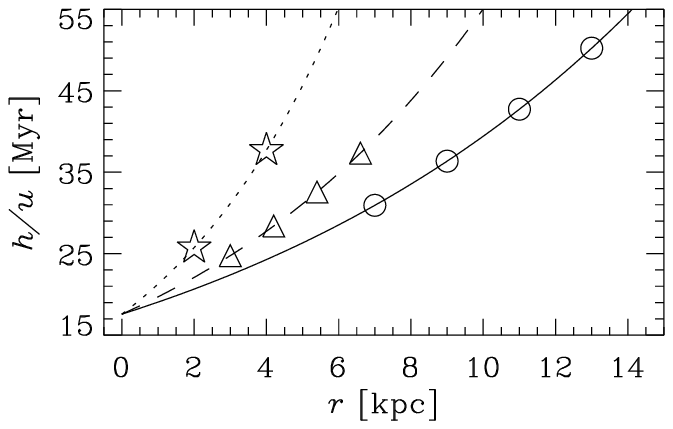}
  \includegraphics[width=0.32\textwidth]{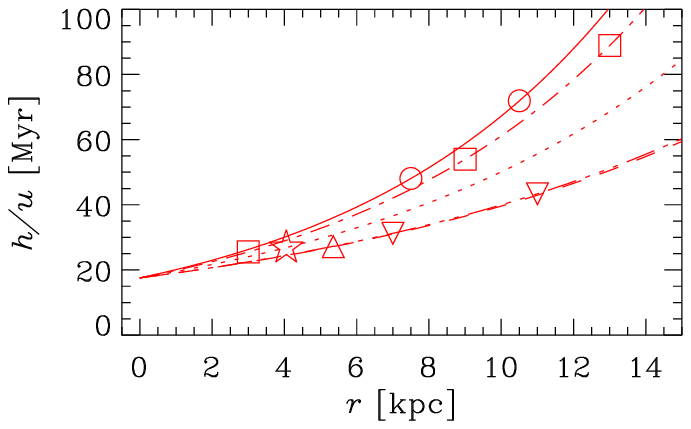}
  \includegraphics[width=0.32\textwidth]{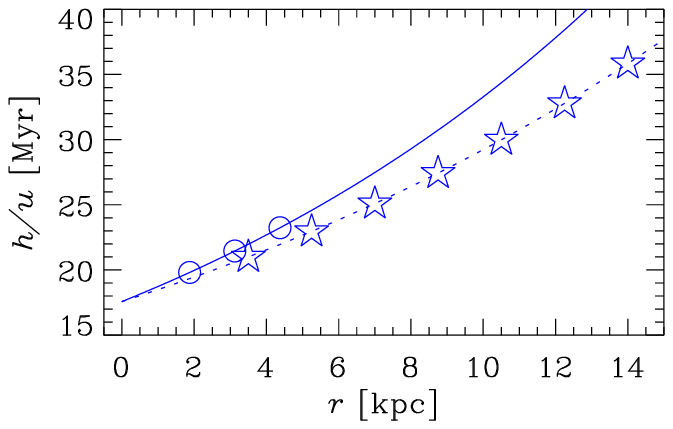}
  \caption{Key fiducial model input for the galaxies in our sample, 
           derived, in part, from observational data of Table~\ref{tab:input_data}.
           \textbf{Column~1:}
           Group~I galaxies
           M31 (circles), 
           M33 (stars), and
           M51 (triangles). 
           \textbf{Column~2:}
           Group~II galaxies
           M81 (circles), 
           NGC~253 (star), 
           NGC~1566 (upward-pointing triangle), 
           NGC~6946 (squares), and 
           IC~342 (downward-pointing triangles).
           \textbf{Column~3:}
           Group~III galaxies
           NGC~1097 (circles) and 
           NGC~1365 (stars).
           \textbf{Top~row:}    Dimensionless outflow velocity $R_U=3HV$ against galaxy radius $r$. 
                                Galaxies for which $3HV$ could not be estimated 
                                due to star formation rate density data being unavailable are not plotted,
                                but are assigned $V=0$.
                                Both group~III galaxies were assigned $V=0$, hence the top right panel is omitted.
           \textbf{Bottom~row:} Ratio of disk semi-thickness to turbulent speed $h/u$ against $r$.
                                The analytical function used for each galaxy has been drawn through the points,
                                with a different line style for each galaxy.
           \label{fig:input}}
\end{figure*}  

\begin{figure*}                       \includegraphics[width=0.32\textwidth]{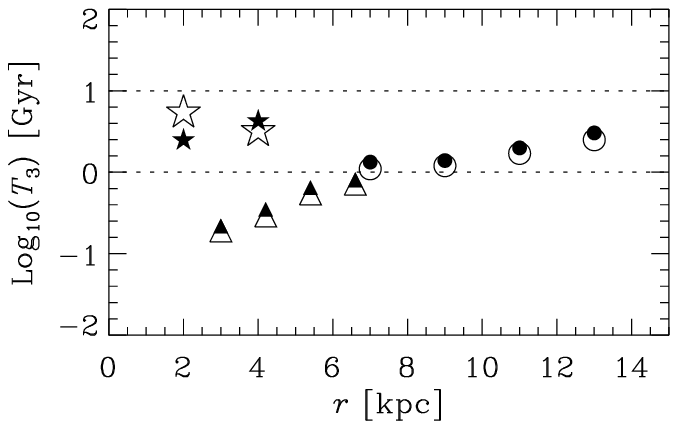}
  \includegraphics[width=0.32\textwidth]{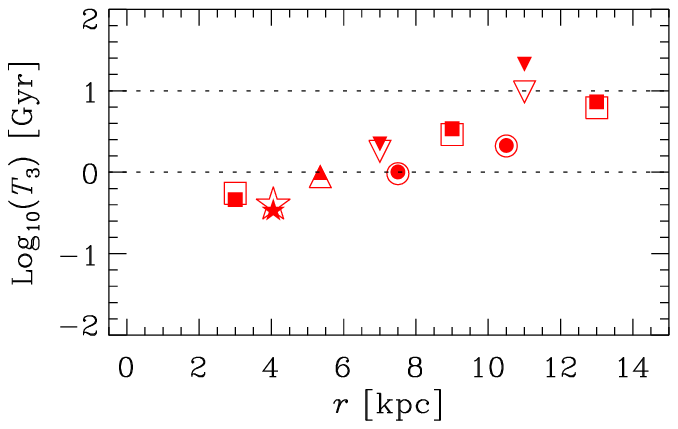}
  \includegraphics[width=0.32\textwidth]{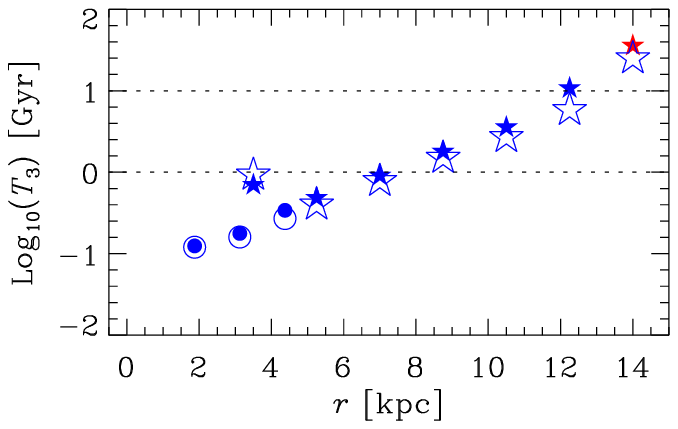}
  \caption{As Figure~\ref{fig:input} but now showing the $10^3$-folding local growth time $T_3$ against $r$.
           Open and closed symbols are for numerical and analytical solutions, respectively.
           For the outermost data point of NGC~1365, the analytically derived value of $T_3<0$, 
           i.e. the field decays (for the analytical but not the numerical solution); 
           its absolute value is plotted as a solid star of a different color.
           Dotted lines show $T_3=1\Gyr$ and $10\Gyr$ for reference.
           \label{fig:growth_time}}
\end{figure*}

\section{Galaxy models}
\label{sec:input}
In this section we derive the dynamo parameters \eqref{dimensionless} for the sample galaxies
using the observational data of Table~\ref{tab:input_data}.
The galactic rotation curves and, hence, the rotation $\Omega(r)$ and shear rates $q(r)$ are reasonably 
well constrained; these parameters are the same as those in \citetalias{Vaneck+15}.

A suitable estimate of the magnitude of the mean vertical velocity at the disk surface, $U\f$, 
can be found in Appendix~B1 of \citetalias{Vaneck+15}. With a typographic error in that paper corrected, 
we have
\begin{equation}
  \label{muz}
  U\f= 0.2\kms h\Sigma_\mathrm{I}^{-1}\Sigma_\mathrm{*}^{1/3}n_\mathrm{h}^{2/3},
\end{equation}
where $h$ is the scale height of the ionized gas layer normalized by $0.5\kpc$, 
$\Sigma_\mathrm{I}$ is the HI surface density in units of $M_\odot\pc^{-2}$,
$\Sigma_\mathrm{*}$ is the surface density of star formation rate in units of $M_\odot\pc^{-2}\Gyr^{-1}$,
and $n_\mathrm{h}$ is the number density of the hot phase normalized by $10^{-3}\cmcube$.

This form of $U\f$ leads to $V=U\f/u$ proportional to $h/u$. Then, since $H\propto h/u$, virtually all 
features of the dynamo solutions depend only on the ratio $h/u$, rather than $h$ and $u$ separately. 
The only variable that depends on $h$ and $u$ differently is $B\eq=u\sqrt{2\pi\Sigma_\mathrm{I}/h}$. The value
of $B\eq$ enters the steady-state magnetic field strength but not the pitch angle
\eqref{psat} and \eqref{pkin} and the kinematic growth rate \eqref{gamma} of the mean magnetic field.

In our fiducial model we assume that the ionized disk is flared, 
with an exponential increase in the disk scale height with galactocentric radius. 
For the Milky Way, 
we adopt the exponential scale length of $10\kpc$,
as in the \ion{H}{1} disk model of \citet{Kalberla+Dedes08}.
We also set $h|_{r=8\kpc}=0.4\kpc$ for the Milky Way, 
consistent with observed values of the warm neutral medium in the solar neighborhood 
\citep{L84} \citep[see also Sect.~VI.2 of][]{Ruzmaikin+88}.
This exponential relation is then rescaled using the $r_{25}$ radius. 
Adopting $r_{25}=16\kpc$ for the Milky Way \citep{Bigiel+Blitz12},
we have
\begin{equation}
  \label{h}
  h= h\f\Exp{r/r\f}
\end{equation}
with $h\f=(0.4\kpc)\exp(-8\kpc/10\kpc)\approx0.18\kpc$ 
and $r\f=10\kpc\,(r_{25}/16\kpc)$.

We adopt a turbulent speed that does not vary within or between galaxies, $u=10\kms=\const$, corresponding
to transonic turbulence in the warm interstellar gas of a temperature about $10^4\K$.
There is some evidence of variations of the turbulent speed with galactocentric radius,
discussed in Section~\ref{sec:velocity_dispersion}, but it is perhaps not conclusive enough.

Figure~\ref{fig:input}
shows the key parameters of the galaxy models obtained from the data of Table~\ref{tab:input_data}.
Columns correspond, from left to right, to galaxy Groups~I--III, as defined in Section~\ref{sec:data}.
The top row presents the radial variation of $R_U=3HV$, a dimensionless quantity that determines the 
importance of the galactic outflow via equations~\eqref{psat}, \eqref{psi} and \eqref{Bsat}).
Note that $3HV\ll1$ in all cases for which the required data is available,
so outflows hardly affect the dynamo action in the model adopted.
For those galaxies where $\Sigma_\mathrm{*}$ is not listed in Table~\ref{tab:input_data},
we adopt $U\f=0$. Given the uncertainties involved in our estimate of the outflow speed,
equation~\eqref{muz}, we return to address the role of outflows in Section~\ref{sec:discussion}.
The bottom row of Figure~\ref{fig:input} shows the radial profile of $h(r)/u$ used in our fiducial model.

For the turbulent correlation time $\tau$ we adopt $14\Myr$, 
which is slightly larger than the value 
of $\tau=l/u\approx10\Myr$ with $l\approx100\pc$ and $u\approx10\kms$ often adopted in galactic dynamo models,
but in the middle of the physically plausible range of 3--$30\Myr$.
This choice leads to a good overall agreement with the observational data on the magnetic pitch angle
while producing positive local kinematic growth rates for all data points.
Restricting $\tau$ to be the same for all galaxies is probably too stringent, 
as $\tau$ is likely to depend on other factors such as the star formation rate and the local sound speed.
However, to keep the model, analysis, and presentation simple, we prefer to adopt a constant $\tau$ 
in our fiducial model, but discuss the implications of relaxing this constraint in
Section~\ref{sec:variable_tau}.

Only dynamo solutions with a positive local growth rate $\gamma$ of equation~\eqref{gamma}
are physically significant as only then the model produces a non-zero mean magnetic field. 
In order to satisfy this constraint for all the data points,
we are forced to adopt $C_\alpha=2$, rather than the standard value of unity.
With $C_\alpha=1$, eight of the $29$ data points have decaying magnetic fields in the 
kinematic regime, $\gamma<0$.
According to equation~\eqref{psat}, the magnetic pitch angle $p$ in the saturated regime is 
insensitive to the value of $C_\alpha$, and numerical solutions confirm this.
A change from $C_\alpha=2$ to $C_\alpha=4$ in the numerical solution does not produce more 
than a few per cent difference in $p$.
Thus, in Regions~(a) and (b) of parameter space, shown in Figure~\ref{fig:pspace},
$\alpha\kin$ is enhanced by a factor of $2$,
but in Region~(c) the upper limit $\alpha\kin=u$ is achieved, corresponding to  $K=1$.
In each case where $\Coriolis>1$, 
the amplitude of $\alpha\kin$ reaches the upper limit $Ku$.
Therefore, for fiducial model parameters, the relevant regions of parameter space are (a) and (c) only.
The amplitude of $\alpha\kin$ reaches the upper limit of $10\kms$ for only five of the $29$ data points,
all at galactocentric distances $r\lesssim4$--$5\kpc$: 
in the innermost annulus of M51, at all radii in NGC~1097,
and in the innermost annulus of NGC~1365.

\begin{centering}
\begin{figure*}                     
  \includegraphics[width=0.33\textwidth]{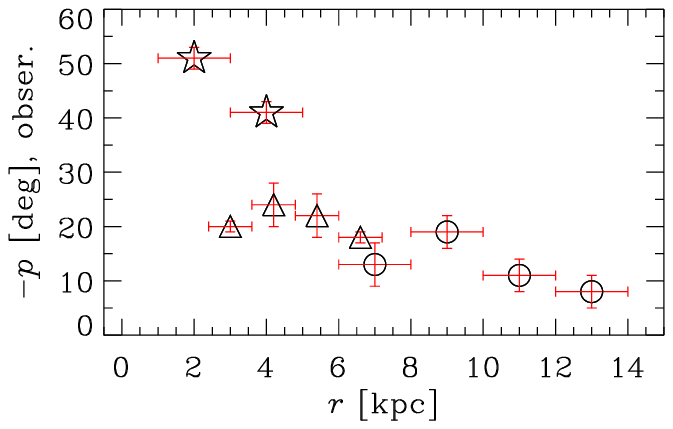}
  \includegraphics[width=0.33\textwidth]{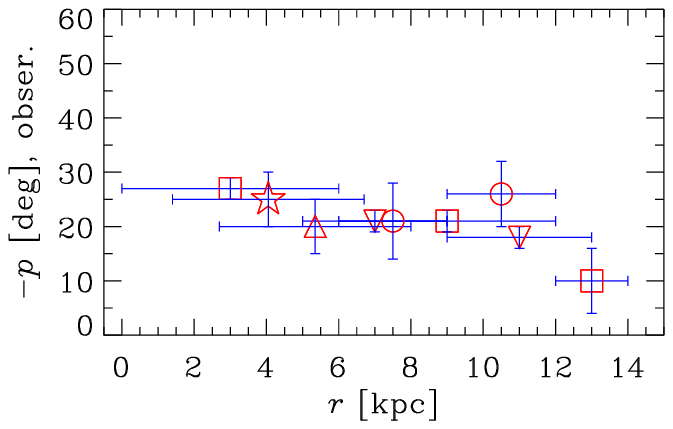}
  \includegraphics[width=0.33\textwidth]{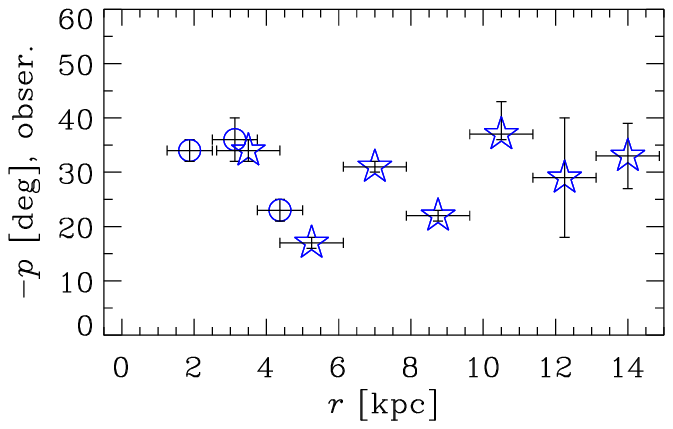}
  \\
  \includegraphics[width=0.33\textwidth]{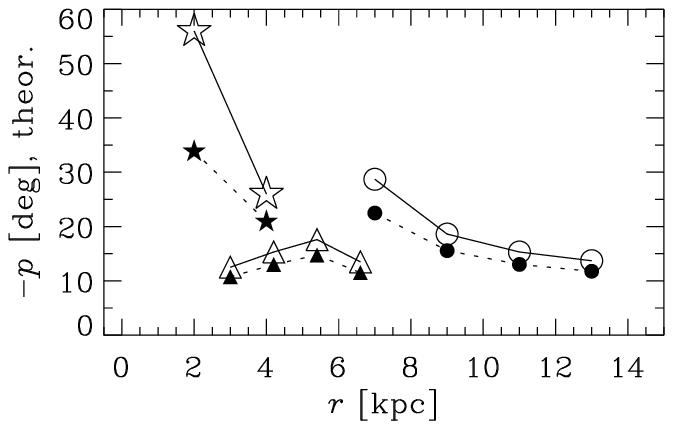}
  \includegraphics[width=0.33\textwidth]{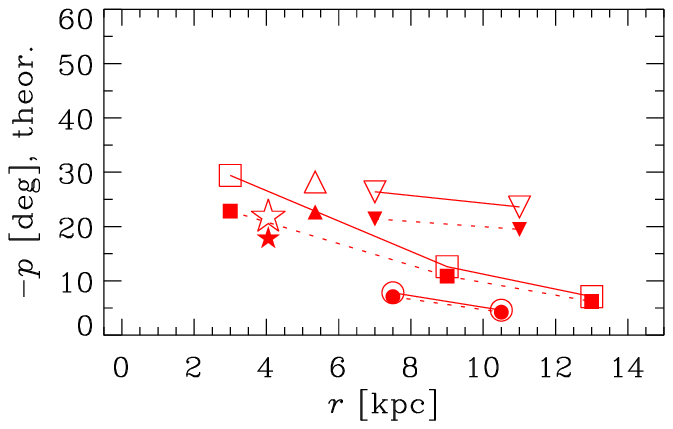}
  \includegraphics[width=0.33\textwidth]{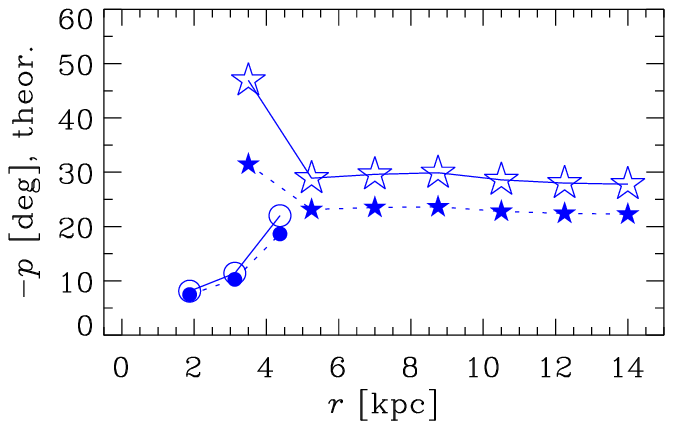}
  \\
  \includegraphics[width=0.33\textwidth] {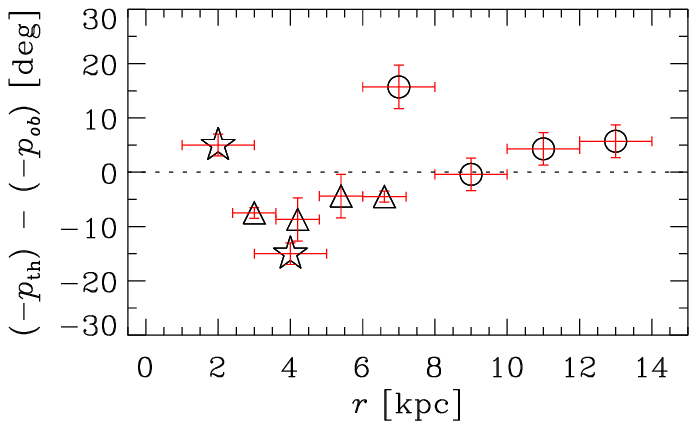}
  \includegraphics[width=0.33\textwidth] {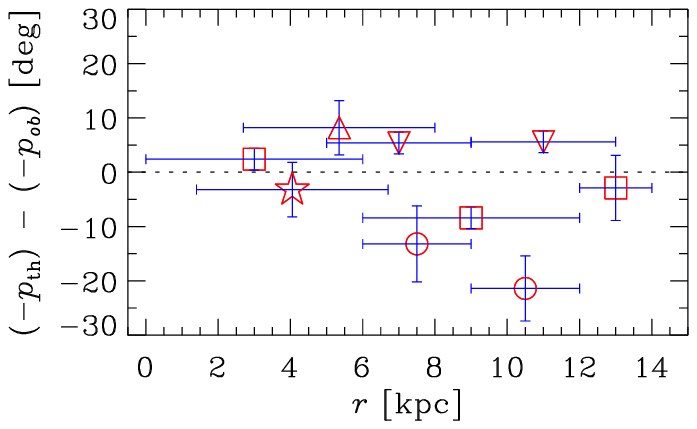}
  \includegraphics[width=0.33\textwidth] {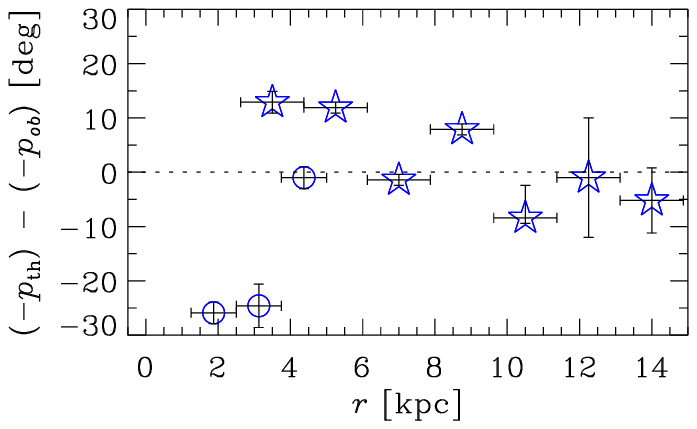}
  \caption{Symbols and columns as in Figure~\ref{fig:growth_time}.
           \textbf{Top~row:}    Pitch angle data.
           \textbf{Middle~row:} Pitch angle model output.
           \textbf{Bottom~row:} Difference between theoretical and observational values of $-p$.
           \label{fig:p}}
\end{figure*}                       
\end{centering}

\begin{centering}
\begin{figure*}                     
  \includegraphics[width=0.32\textwidth]{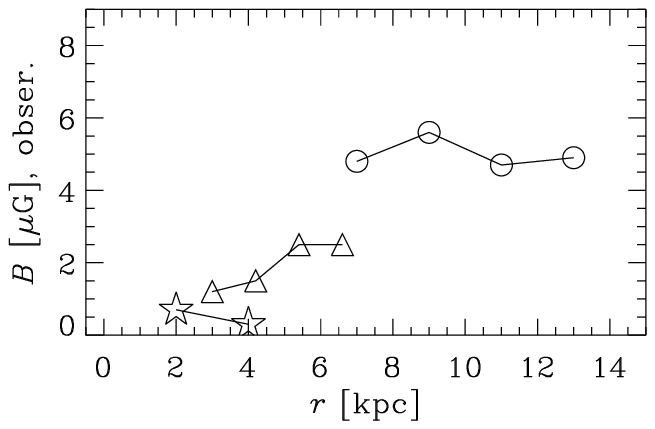}
  \includegraphics[width=0.32\textwidth]{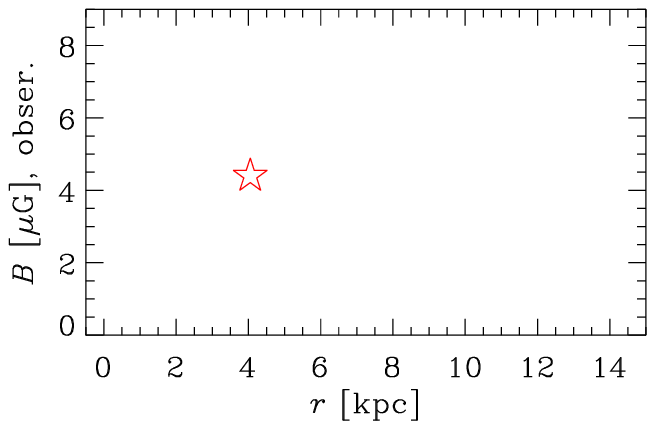}
  \includegraphics[width=0.32\textwidth]{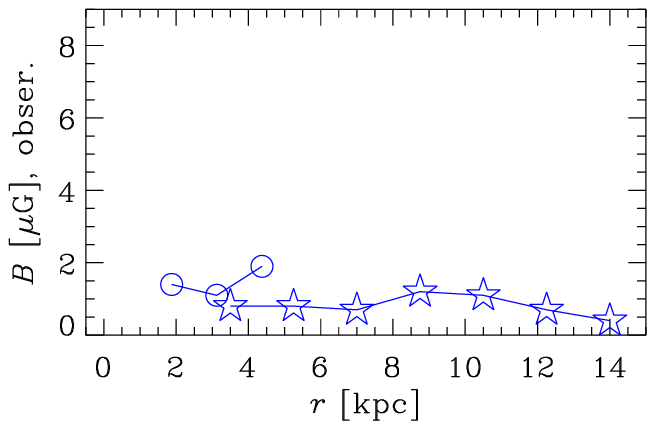}
  \\
  \includegraphics[width=0.32\textwidth]{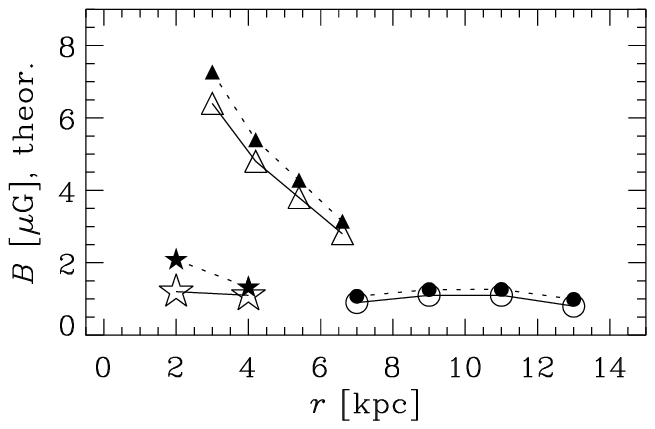}
  \includegraphics[width=0.32\textwidth]{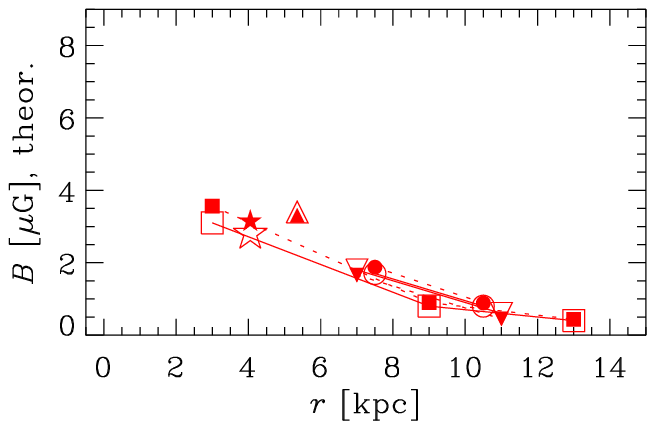}
  \includegraphics[width=0.32\textwidth]{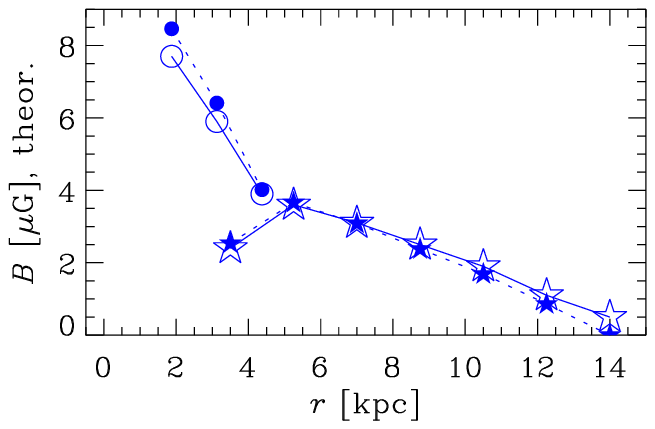}
  \\
  \includegraphics[width=0.32\textwidth] {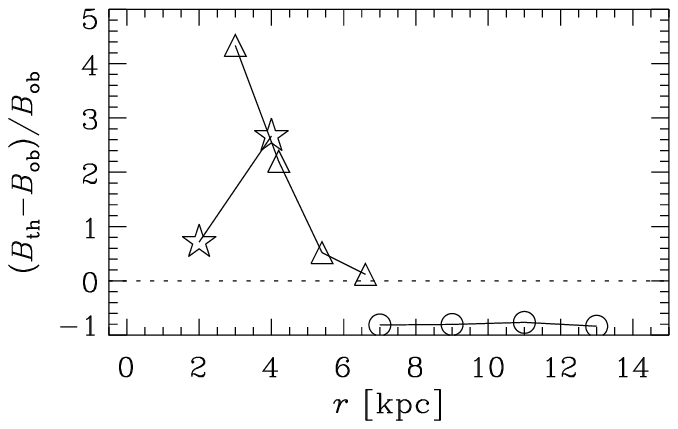}
  \includegraphics[width=0.32\textwidth] {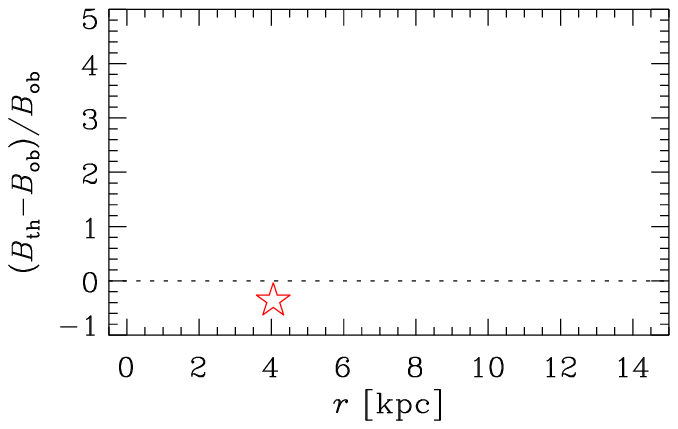}
  \includegraphics[width=0.32\textwidth] {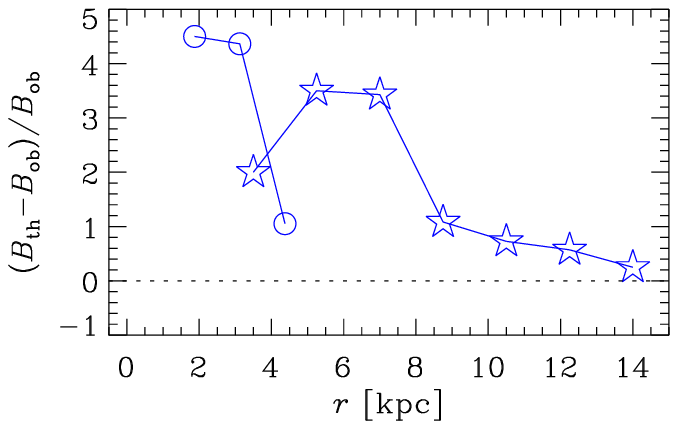}
  \caption{Symbols and columns as in Figures~\ref{fig:growth_time} and \ref{fig:p}.
           \textbf{Top~row:}    Mean magnetic field strength data.
                                Those galaxies for which data are unavailable are not plotted.
           \textbf{Middle~row:} Mean magnetic field strength model output.
           \textbf{Bottom~row:} Difference between theoretical and observational values of $\mean{B}$,
                                normalized with respect to observed values,
                                for those galaxies for which data are available.
           \label{fig:B}}
\end{figure*}                       
\end{centering}

\begin{centering}
\begin{figure*}                     
  \includegraphics[width=0.32\textwidth]{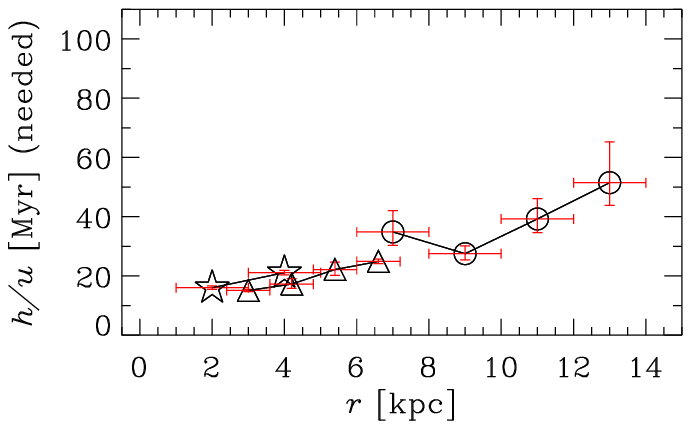}
  \includegraphics[width=0.32\textwidth]{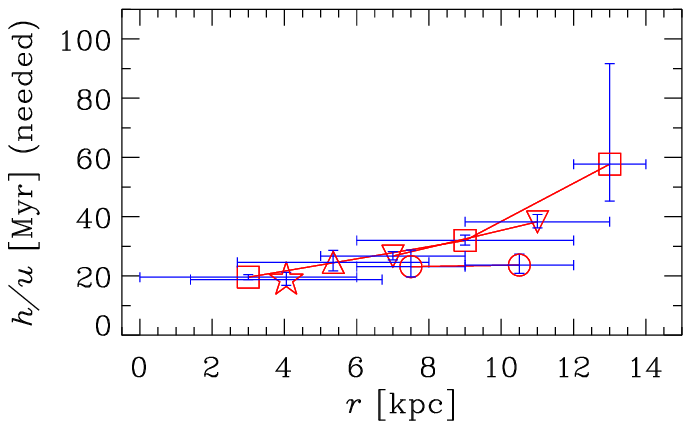}
  \includegraphics[width=0.32\textwidth]{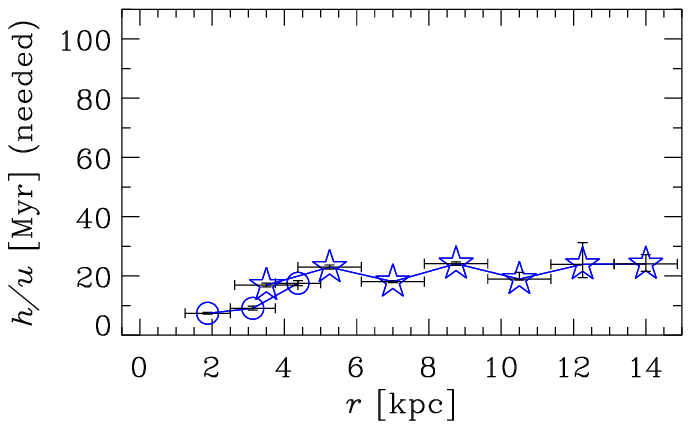}
  \\
  \includegraphics[width=0.32\textwidth]{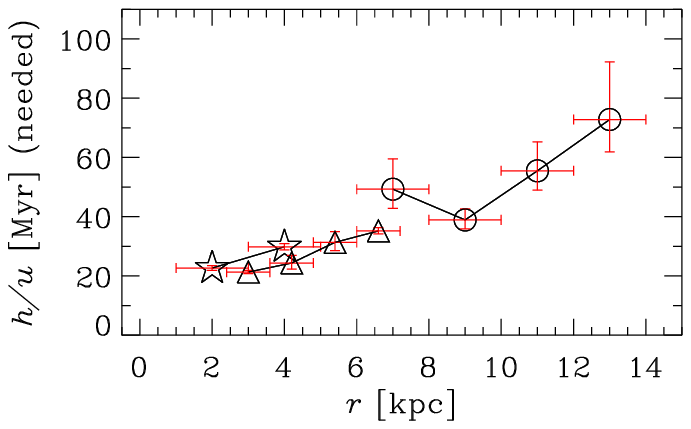}
  \includegraphics[width=0.32\textwidth]{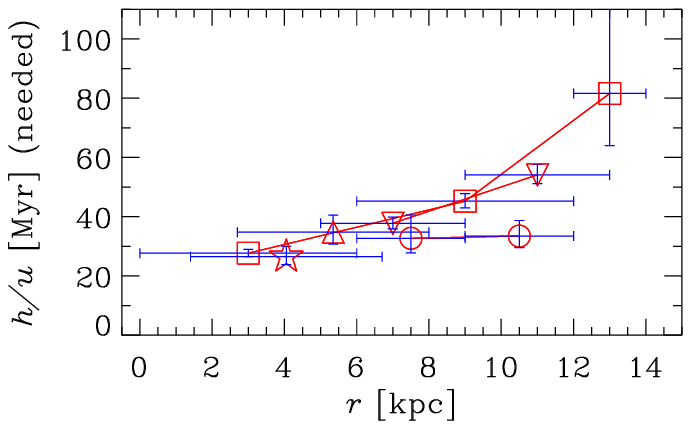}
  \includegraphics[width=0.32\textwidth]{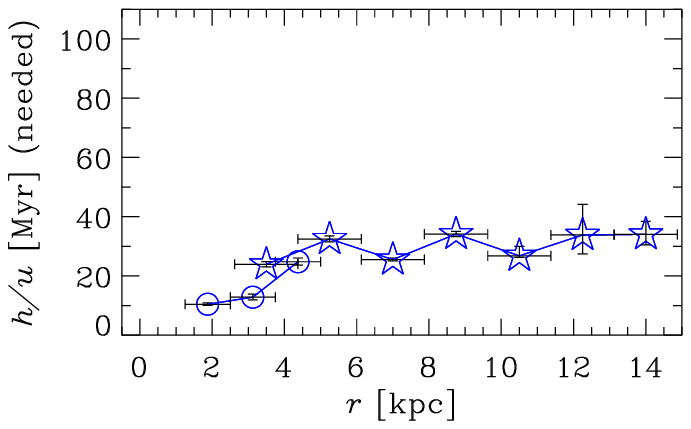}
  \\
  \includegraphics[width=0.32\textwidth]{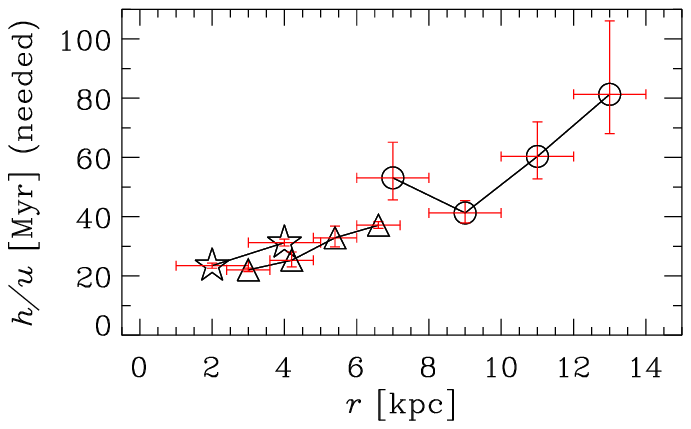}
  \includegraphics[width=0.32\textwidth]{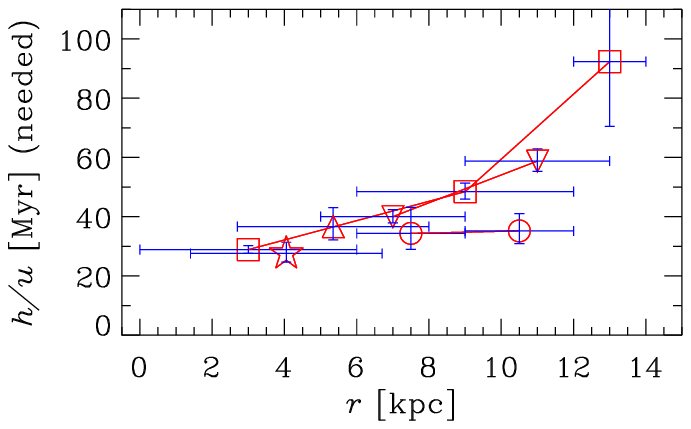}
  \includegraphics[width=0.32\textwidth]{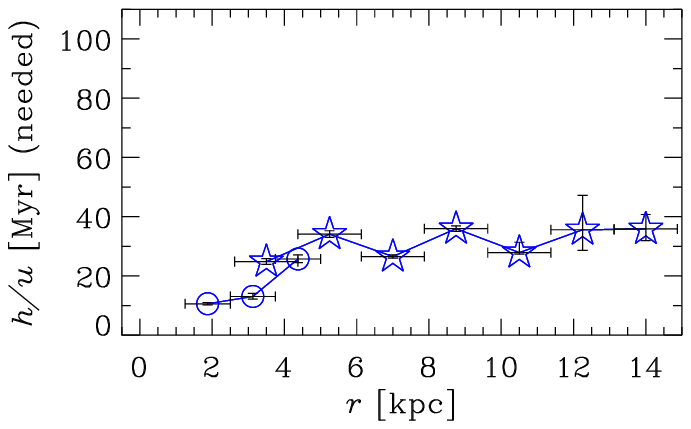}
  \caption{Values of the ratio $h/u$ needed to obtain agreement with pitch angle data, for the analytical solution.
           Symbols and columns as in Figure~\ref{fig:input}.
           \textbf{Top row:}    $\tau=10\Myr$, $V=0$. 
           \textbf{Middle row:} $\tau=20\Myr$, $V=0$. 
           \textbf{Bottom row:} $\tau=20\Myr$, $V=0.1$. 
           Vertical error bars are estimated by substituting the observed $p$ by the values at the ends of the $p$ error bars.
           The upper error bar on the outermost data point of NGC~6946 (which exceeds the plotting range) extends to $130\Myr$ and $158\Myr$,
           in the middle and bottom rows, respectively.
           These values of $h/u$ can be compared with those used as input to the model, 
           shown in the bottom row of Figure~\ref{fig:input}.
           \label{fig:hu}}
\end{figure*}                       
\end{centering}

\section{Results}
\label{sec:output}
\subsection{Growth time of the mean magnetic field}
\label{sec:T3}
It is often claimed that the mean-field galactic dynamos generate magnetic fields at a time scale exceeding
$1\Gyr$, too large to explain magnetic fields suggested to be present in galaxies at $z>1$ (age of order $5\Gyr$)
\citep{BMLKD-Z08,Bernet+13}. 
We present in Figure~\ref{fig:growth_time} the $10^3$-folding growth times $T_3$, 
for numerical (open symbols) and analytical (filled symbols) dynamo solutions, 
as a function of galaxy radius for each of the galaxy groups I--III.
The amplification factor of $10^3$ is chosen because the fluctuation dynamo action is expected
to produce an effective seed magnetic field for the galactic mean-field dynamos of order
$10^{-3}\mkG$ \citep[Section VII.14 in][]{Ruzmaikin+88,Shukurov07}. 
In almost all cases, $T_3<10\Gyr$ and $T_3<1\Gyr$ at $r\lesssim8\kpc$.
Furthermore, our model implicitly assumes that galaxy properties such as $\Omega$ and $q$
have not evolved during the course of dynamo action;
this assumption is probably not important for the saturated state, 
but could be important for estimates of the growth rate.

\subsection{Magnetic pitch angle}
\label{sec:pitch}
Both the observational estimates of and predictions of the dynamo model for the magnetic pitch angle
in the sample galaxies are shown in Figure~\ref{fig:p}.
As before, open symbols represent numerical solutions while filled symbols are for the analytical solutions,
both in the saturated dynamo. 
Numerical solutions are obtained for $R_\kappa=0.3$,
the value obtained from MHD simulations by \citet{Mitra+10};
$p$ is hardly sensitive to $R_\kappa$.

The top row of Figure~\ref{fig:p} shows observational data for $p$ as a function of galactocentric radius $r$,
with uncertainties in $p$ taken from the original works cited by \citetalias{Vaneck+15}.
The data points are referred to the middle of the radial ranges for which they are obtained, 
and those radial ranges are shown with horizontal bars.
In the middle row, the pitch angles from the dynamo solutions are shown,
with open (filled) symbols denoting numerical (analytical) solutions.
The bottom row shows the difference between the theoretical and observed pitch angles.

The pitch angles from the fiducial model are given in Column~5 of Table~\ref{tab:B_data}, 
and the observational values are in Column~3.
We cannot expect to achieve agreement between the model and observations at the level
of parametric statistical tests because of the heterogeneous nature of the data and the deliberately
simplified form of the dynamo field model (see Section~\ref{PRGDM}). Nevertheless, to provide 
a feeling for the relative quality of the fits, we quantify 
the agreement between various versions of the model and observations in terms of their residual
difference, $\chi^2$. We obtain a reduced $\chisq$ of $\chisqred=\chisq/\nu=18$, $7$, and $51$ 
for numerical solutions of group~I, II, and III galaxies, respectively,
where $\nu$ is equal to the difference between the number of data points 
and number of free parameters.

Kendall's rank correlation test can be used to assess whether the model reproduces the trends in the
data at a statistically significant level. Having accepted that the model may systematically underestimate
or overestimate the magnetic pitch angles, this test allows us to find out if the difference between the
model and observations is systematic and consistent (which would speak in favor of the model) or 
just random.
Let $N$ be the number of galaxies within a group, and $n_i$ the number of data points in galaxy $i$.
The number of galaxy pairs within a group is  $N_1=N(N-1)/2$,
and $N_2=\sum_{i=1}^N n_i(n_i-1)/2$ is the number of intra-galaxy pitch angle pairs within a group.
Then $N_1\cor\le N_1$ is the number of galaxy pairs
that are correctly ordered by the model according to mean value of $p$;
that is the number of concordant pairs, as opposed to discordant pairs. 
In this notation, Kendall's $\tau=2N_1\cor/N_1-1$.
Likewise, $N_2\cor\le N_2$ is the number of intra-galaxy pairs 
of the $p$ values that are correctly ordered.

We thus find that $N_1=3$, $10$, and $1$, 
and $N_2=13$, $5$, and $24$ for groups I, II, and III, respectively.
For our model, we obtain $N_1\cor=2$, $5$, and $0$, and $N_2\cor=10$, $4$, and $11$.
The probability $P_1$ of obtaining $N_1\cor\ge N\f$ (where the values of $N\f$ that now concern us are $2$, $5$, or $0$), 
given a purely random ordering (the null hypothesis),
can be computed using inversion theory and the result depends on the Mahonian sequence of numbers.\footnote{See 
\citet{Kendall38} and http://oeis.org/A008302. 
For example, for $N=3$, we find $N_1=3\cdot2/2=3$ pairs.
The probabilities for obtaining $0$, $1$, $2$, or $3$ correct orderings by chance 
are $1/6$, $1/3$, $1/3$, and $1/6$, respectively.
These are computed by dividing the appropriate part of the Mahonian sequence $\{1,2,2,1\}$ by $N!=6$.
The probability $P_1$ for obtaining $2$ or greater correct orderings by chance is then $1/3+1/6=1/2$.
To obtain $P_2$, the extra step of summing the probabilities of all relevant permutations is needed.}
We obtain $P_1=0.5$, $0.59$, and $1$,
which tells us that our model is no better than the null hypothesis 
at predicting the relative ordering of of the pitch angles.
We can also define $P_2$ in the analogous way, using $N_2$ and $N_2\cor$. 
We obtain $P_2=0.08$, $0.21$, and $0.66$.
This provides mild to moderate evidence that the model is better 
than a uniform distribution over permutations at correctly
ordering the intra-galaxy pairs of pitch angles for groups~I and II,
but not for group~III (two barred galaxies).

\subsection{Strength of the large-scale magnetic field}
\label{sec:B}
Figure~\ref{fig:B} shows the
strength of the mean magnetic field $\mean{B}$ 
obtained from observations and the dynamo model, as well as the relative difference between the two in the bottom row.
The observational estimates in Groups~I and III are obtained from the amplitude of the $m=0$ mode in each radial bin.
In Group~II, however, radially binned field strengths are available for only one galaxy, NGC~253.
Uncertainties on the observed values of $\mean{B}$ were not readily available.
The dynamo model typically overestimates the field strength, but only by a factor of about $2$ on average, 
with the notable exceptions of M31 where the field strength is underestimated by a factor of 
about five, 
and NGC~253 where the theoretical estimate is in reasonable agreement with the observed value.
The radial variation in field strength is in general poorly reproduced by the model,
but for the galaxy M31 the variation is negligible, in agreement with observations.
Unlike $p$, the field strength
obtained from the dynamo model is sensitive to a larger number of parameters,
such as $R_\kappa$, $C_\alpha$, and $\Strouhal$.
We choose $R_\kappa=0.3$ \citep{Mitra+10} rather than $1$ specifically to improve 
the agreement with observations.
Moreover, 
$\mean{B}$ depends on the values of $u$ and $h$ individually (through $B\eq$) rather than just their ratio,
as well as on our rather crude assumption that the surface density of ionized gas is equal to that of HI gas.
Equally important is the fact that the estimates of $\mean{B}$ from Faraday rotation are subject to various
systematic uncertainties, such as the effects of correlations between thermal electron density and magnetic
fields \citep{Beck+03}. With allowance for all these uncertainties, we conclude that the agreement
between theory and observations for the large-scale magnetic fields strength is reasonable, even if further
improvements are clearly desirable on both theoretical and observational sides.

\section{Model variations and refinements}
\label{sec:variations}
The experience gained in this paper helps to understand how galactic dynamo models
should be developed to achieve better agreement with observations; these aspects of
dynamo models are discussed in this section. We have tentatively explored some such
directions, but their thorough analysis extends beyond the framework of this paper.

\subsection{The role of disk flaring}
\label{sec:constant_h}
To clarify the importance of the disk flaring, we consider a dynamo
model 
with a flat disk,
$h=0.4\kpc=\const$, but other parameters unchanged.
Results for $p$ are shown in Column 9 of Table~\ref{tab:B_data}, 
and yield $\chisqred=71$, $18$, and $119$, 
$N_1\cor=2$, $1$, and $0$, and $N_2\cor=2$, $1$, and $14$ 
for the Groups~I, II, and III, respectively.
This is worse agreement than with a flared disk, and varying $\tau$ does not help.

The radial variation of $h/u=H\tau$ that would fit the observed magnetic pitch angles can be obtained
from equation~\eqref{psat},
as shown in Figure~\ref{fig:hu} as derived from the analytical solution 
for $\tau=10\Myr$ and $V=0$ (top row), $\tau=20\Myr$ and $V=0$ (middle row), 
and $\tau=20\Myr$ and $V=0.1$ (bottom row).
An increase in the required $h/u$ with $r$ is evident, 
both within and between galaxies, for Groups~I and II.
For the barred galaxies of Group~III, the required $h/u$ is much flatter with respect to $r$.
The dependence of $h$ on $r$ shown in Fig.~\ref{fig:hu} is comparable
with that used in the fiducial dynamo model and shown in Fig.~\ref{fig:input}. 
Since the radial variations in both the scale height 
of the ionized layer and the turbulent velocities
are not known confidently from either observations or theory, 
we tend to consider the results presented in Fig.~\ref{fig:hu} as a prediction.

\subsection{Turbulence correlation time}
\label{sec:variable_tau}
The turbulence correlation time $\tau$ is likely to vary from galaxy to galaxy
and within a given galaxy, for instance because it can depend on the star formation rate and detailed
dynamics of interstellar turbulence \citep{Shukurov07}.
Thus, we consider dynamo models with a value of $\tau$, 
for the sake of simplicity chosen to be constant for each galaxy,
that minimizes $\chi^2$ 
(and $\tau$ considered as an additional parameter in the calculation of $\chisqred$) 
for a given galaxy:
$\tau=10\Myr$ for M31, $14\Myr$ for M33, $20\Myr$ for M51, $16\Myr$ for NGC~253, 
$19\Myr$ for NGC~1097, $12\Myr$ for NGC~1365, $10\Myr$ for NGC~1566, $14\Myr$ for NGC~6946, and $11\Myr$ for IC~342.
The optimal values of $\tau$ are within a factor of two of each other and comparable 
with the existing estimates \citep[e.g.,][]{Shukurov07}. 
Only for one galaxy, M81, the required value exceeds $30\Myr$, so we set $\tau$ to $30\Myr$ for it. 
The resulting magnetic pitch angles in the saturated state 
of the dynamo are shown in Column 8 of Table~\ref{tab:B_data}.
We obtain $\chisqred=11$, $7$, and $37$, and $N_1\cor=3$, $7$, and $0$, 
with associated null probabilities $P_1=0.17$, $0.24$, and $1$ 
for the galaxies in Groups~I, II, and III, respectively 
($N_2\cor$ is unchanged from the constant-$\tau$ model).
Thus, the dynamo model agrees with the data somewhat better 
when $\tau$ varies between galaxies in Group~I, 
but only marginally in Groups~II and III. 
It is equally likely that $\tau$ varies within galaxies as well.
We do not pursue this possibility further 
before more definite estimates of the turbulence correlation time are available. 
However, our results provide strong indication that this variation 
is at least within a factor of several within and between spiral galaxies.

\subsection{Turbulent velocity}
\label{sec:velocity_dispersion}
Magnetic pitch angle depends strongly on the turbulent velocity,
$\tan p\propto u^2$ from \eqref{psat} or \eqref{psatV0},
neglecting the outflow, $U_0=0$.
Observations of nonthermal \ion{H}{1} broadening suggest, 
for the one-dimensional velocity dispersion
in the radial ranges considered here,
$\sigma=(12$, $12$, $12$, $11)\kms$ in M31 \citep[][C.~Carignan, private communication]{Chemin+09},
$\sigma=(10$, $9)\kms$ in M33 (C.~Carignan, private communication),
and $\sigma=(22$, $22$, $21$, $20)\kms$ in M51 \citep{Tamburro+09}.
The corresponding values of the three-dimensional turbulent speed $u$ 
are $\sqrt3$ times larger if the turbulence is isotropic. 
However, we used a constant turbulent speed of $u=10\kms$ in the dynamo model.
Larger values of $u$ could be accommodated in the model by increasing $h$ proportionately,
while differences in $u$ between galaxies could potentially improve agreement with observations.
However, as data for $\sigma$ is not immediately available for all the galaxies of our sample,
and as inferring $u$ from $\sigma$ may involve additional subtleties, 
we leave such an approach for future work.

\section{Discussion}
\label{sec:discussion}
We are left, then, with a simple dynamo model, which 
provides reasonable agreement with observational estimates of the pitch angle and strength of 
the large-scale magnetic field, and is sufficiently flexible to incorporate various refinements. 
It is notable that the fiducial model, based on the `standard' estimates
of the galactic parameters, needs to be refined, 
for example by using variable turbulence correlation time,
in order to achieve a better agreement. 
Together with this observation, our results strongly suggest 
that the ionized galactic disks are flared 
and provides an estimate of the flaring rate implied 
by magnetic field observations in nearby spiral galaxies. 
In this section we discuss the implications of our results for galactic dynamos.

\subsection{The role of galactic outflows}
\label{sec:outflows}
Despite growing evidence of the importance of galactic outflows (winds and fountains) 
for many aspects of physics of galaxies \citep{VCB-H05,Putman+12}, 
estimates of the outflow speed, and its dependence on the galactic parameters, 
remain uncertain. 
The dynamo model used here relies on the order-of-magnitude estimate 
of the outflow speed \eqref{muz}. 
We could estimate an outflow speed for only five
of the ten galaxies in our sample (M31, M33, M51, NGC~253, and NGC~6946),
but the gas surface density and/or star formation rate surface density were not readily available
for the remaining galaxies.
Equation~\eqref{muz} predicts that the outflows are rather weak, $3HV\lesssim1$ to affect the magnetic pitch angle
to any significant degree. 
As shown in Column~10 of Table~\ref{tab:B_data} adopting $U_0=0$ affects the
magnetic pitch angles insignificantly.

To allow for the uncertainty of Eq.~\eqref{muz}, 
we also considered models with the outflow speed 
ten times larger than in \eqref{muz} but with the same dependence on galactic parameters.
The resulting values of the magnetic pitch angle shown in Column~11 of Table~\ref{tab:B_data}, 
are by about 10 per cent larger than in the fiducial model,
up to a maximum of about $40$ per cent for the outermost data point in NGC~6946.
For the reader's convenience, the analytical solution of the
dynamo equations with outflows neglected, $U_0=0$, are presented in Appendix~\ref{sec:V0}.
Galactic outflows play an important role in the large-scale dynamo action as they help to
remove magnetic helicity from the dynamo-active region among other effects 
\citep{Shukurov+06,Sur+07b,Chamandy+14b,Chamandy+15}. 
Their effects on observable parameters of galactic magnetic fields clearly require further analysis.

\begin{figure}
  \begin{centering} \includegraphics[width=0.8\columnwidth]{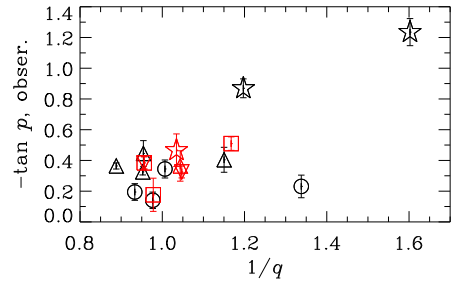}
    \caption{Tangent of observed magnetic pitch angle, $-\tan p$, plotted against inverse shear parameter $1/q$,
             for the galaxies M31, M33, M51, NGC~253, NGC~1566, NGC~6946, and IC~342,
             with symbols and colors the same as in Figures~\ref{fig:input}--\ref{fig:B}.
             A correlation is found to exist between these quantities.       \label{fig:spiral}
            }            
  \end{centering}
\end{figure}

\subsection{The role of spiral arms}
\citetalias{Vaneck+15} found that the pitch angle of the regular magnetic field 
is correlated with that of the spiral arms and suggested that compression
of gas and magnetic field in the arms can explain the correlation.
In addition, equation~\eqref{psat} shows that $|p|$ is anticorrelated with $q$,
the velocity shear rate due to the galactic differential rotation. 
Figure~\ref{fig:spiral} shows the dependence of $-\tan p$ on $1/q$ 
from the data in Tables~\ref{tab:B_data} and \ref{tab:input_data} 
for the galaxies in Groups~I and II, 
excluding M81 for reasons discussed in Section~\ref{sec:M81}.
Pearson's correlation coefficient is $r\Pearson=0.73$ 
(with the adjusted value $r\adj=r\Pearson[1+(1-r\Pearson^2)/(2n)]=0.74$,
where $n=17$ is the number of data points).
This leads to a student-t value of $t=r\Pearson\sqrt{(n-2)/(1-r\Pearson^2)}=4.2$.
The probability of attaining $t\ge4.2$ if the variables are uncorrelated is less than $0.001$.
The dependence of the magnetic pitch angle of an axially symmetric magnetic field 
on the velocity shear arises naturally in the dynamo theory 
because differential rotation is an integral part of the $\alpha$-$\Omega$ dynamo.\footnote{However, 
the magnitude of the shear is expected to be different within spiral arms and between them. 
This effect needs to be included into discussions of the effects of the spiral pattern on galactic magnetic field.}

We also note that a strong negative correlation between 
the pitch angle of spiral arms $|p\arm|$ and the shear rate $q$ 
is found from both observations and simulations \citep{Seigar+05,Seigar+06,Grand+13}.
Thus, at least a part of the apparent correlation between the pitch angles 
of the large-scale magnetic field and spiral arms 
can be due to their dependence on $q$ rather than any causal connection.
However, the two pitch angles not only
depend similarly on the rotational velocity shear $q$ but are also close to each other in magnitude \citepalias{Vaneck+15}. 
Since the two quantities depend on rather different galactic parameters apart from $q$, 
this suggests a causal connection between them.
Clearly, there is a need to further clarify the relationship between the two types of pitch angle,
but this is beyond the scope of the present work.
\subsection{Individual galaxies}
\subsubsection{M33}
\label{sec:M33}
The large pitch angles of M33 have been a challenge to explain using standard dynamo theory \citep{Tabatabaei+08}.
We have shown that such large $|p|$ can indeed be attained
in a dynamo model.
However, though the model correctly produces a $|p|$ that decreases with radius,
it does not do an adequate job of fitting both data points simultaneously,
even allowing for adjustments to $\tau$.
The innermost data point is for the radial bin $1$--$3\kpc$,
where our model, which assumes a thin disk, $h\ll r$, is less reliable.
Another possibility is that accretion of gas with a velocity of a few $\!\kms$, 
as observed for some galaxies \citep{Schmidt+16},
could affect $p$ at small radius in the saturated regime \citep[see Fig.~4 of][]{Moss+00}.

\subsubsection{M81}
\label{sec:M81}
The observed $|p|$ of M81 are much larger than those found from the model
(though because of large formal uncertainties in the M81 pitch angle data,
$\chisqred$ does not improve significantly if M81 is excluded).
M81 is the only galaxy where the large-scale magnetic field may be predominantly
non-axisymmetric \citep{Krause+89,Sokoloff+92,Fletcher10}, 
in which case our model would not be expected to produce good agreement for this galaxy.
However, the observations are rather old, and their interpretation relied on incomplete procedures, 
so that the question of the global magnetic field structure in this galaxy remains open.

\subsubsection{NGC~1097 and NGC~1365}
\label{sec:barred}
The dynamo model used here is axially symmetric, 
so we do not expect it to be accurate in the case
of the barred galaxies NGC~1097 and NGC~1365 in the sample. 
It is reassuring that the dynamo model performs better in the case of Group~I and II galaxies. 
Since deviations from axial symmetry are stronger in the bar region, 
it is understandable that the disagreement between the axisymmetric dynamo model
and observations is greater at small galactocentric distances in the two barred galaxies.
In NGC~1097, the pitch angle for the outermost radial range 
is reproduced by the model to within the observational uncertainty.
Likewise, the solution for the mean magnetic field strength 
in this annulus is within a factor of two of the observed value.
In NGC~1365, the model magnetic pitch angles are consistent with the data in four outer rings.
Numerical dynamo models designed specifically for barred galaxies have been developed by 
\citet{Moss+01,Beck+05,Moss+07,Kulesza-zydzik+10,Kulpa-dybel+11}.

\subsection{The accuracy of magnetic pitch angle estimates}
Magnetic pitch angles obtained from fitting azimuthal Fourier modes 
to the polarization angles,
used to obtain magnetic pitch angles given in Column~3 of Table~\ref{tab:B_data} 
have very small formal uncertainties, 
but their systematic uncertainties can be significantly larger, e.g., those associated
with the variation in the parameters of the magneto-ionic medium along the line of sight, 
relatively crude allowance for depolarization effects, etc. 
Having this in mind, the agreement of the simple dynamo model
used here with observations is encouraging.
For the galaxies M31, M51, NGC~253, NGC~1566, NGC~6946, and IC~342,
our model does a fairly reasonable job of reproducing magnetic pitch angles. 
Allowing the turbulence correlation time $\tau$
to vary between galaxies within a reasonable range $10$--$20\Myr$, 
but still restricted to remain constant within each galaxy,
improves the agreement further.

\subsection{Uncertainty in model parameters}
Although the values of $h$, $\tau$, and $u$ used in our models are in line with typical estimates,
there is still considerable uncertainty in these parameters.
For example,
estimates of $\tau$ from numerical simulations of supernova-driven turbulence
in a stratified medium with rotation and shear yield $\tau\approx2$--$4\Myr$ 
\citep{Gressel+08a,Brandenburg+13}.
Such simulations also suggest $u\approx20$--$30\kms$ \citep{Gressel10,Gent+13a}.
If such modifications to $\tau$ and $u$ are made in the model,
they tend to offset each other as far as the pitch angle is concerned 
(equations~\eqref{psat} and \eqref{psatV0}).
However, the local growth rate is also affected,
and becomes negative for large values of $u$
(equations~\eqref{gamma} and \eqref{gammaV0}).
Likewise, the parameters 
$h$, $\tau$, $u$, $\Strouhal$ and $R_\kappa$ 
have important effects on the field strength 
(equations~\eqref{Bsat} and \eqref{BsatV0}).
Clearly, more work on constraining these parameters 
using observations, simulations, and modeling is needed.

\subsection{Promising refinements of galactic dynamo models}\label{PRGDM}
There are several factors that have not been included in the dynamo model used here.
We consider dynamo action in a thin gas layer in vacuum. 
Disk--halo connections undoubtedly affect magnetic fields in the disk 
but their role still remains to be fully understood.
More generally, considering the full tensorial structure of 
the turbulent transport coefficients and allowing for vertical gradients
of the galaxy parameters leads to additional terms that may be important
in equations~\eqref{dBrdt} and \eqref{dBpdt}. 
For example, 
\[
  \frac{\del\mbi}{\del t}= \ldots -\frac{\del}{\del z}(\gamma_z\mbi) +\frac{\del\eta}{\del z}\frac{\del\mbi}{\del z},
\]
where $i=r$ or $\phi$, dots represent the other terms, 
and the diamagnetic pumping velocity $\gamma_z$ 
can be estimated analytically \citep[e.g.][]{Brandenburg+Subramanian05a}.
Diamagnetic pumping toward the midplane ($\gamma_z<0$) is expected to occur
when $\del u/\del z>0$,
and this effect has been measured in simulations \citep{Gressel+08b,Gressel10}.
Conveniently, this term has the same form as the term involving $\muz$,
and the combination could be crudely approximated by replacing $\muz$ 
in equations~\eqref{dBrdt} and \eqref{dBpdt} by an effective velocity $\muztilde=\muz+\gamma_z$.
The damaging effect of a strong outflow on the dynamo would thus be suppressed
(while the full $\muz$ would presumably still contribute to the advective flux in equation~\eqref{dalp_mdt}).
The advection-type term involving $\del\eta/\del z$ could likewise be important.
It has previously been found, using mean-field models, 
that such terms can play significant roles in the galactic context
\citep{Brandenburg+95,Gabov+96,Shukurov96}.
Including these effects would thus be an important step
toward more realistic modeling.

In fact, it has been suggested that a possible non-linear quenching of $\gamma_z$
could, if present, 
play an equally important role to that of $\alpha$ in the saturation of the dynamo 
\citep{Gressel10,Gressel+13a,Bendre+15}.
Consider for illustration a scenario where quenching of $\gamma_z$ is the dominant saturation mechanism.
Then the value of the effective velocity $\muztilde$ at saturation would be approximately equal 
to that which produces a critical dynamo for $\alpha=\alpha\kin$.
Since large (effective) outflows imply large pitch angles (Sect.~\ref{sec:solution_space}),
this mechanism could potentially produce pitch angles in the saturated state larger
than those in the kinematic regime \citep{Elstner+09}.
Such a scenario requires further theoretical justification,
but could potentially serve as an alternative model to be tested against observations.

Any deviations from axial symmetry in either the host galaxy or the dynamo action 
are also neglected in our treatment.
As a result, the dynamo model employed provides just the lowest-order approximation to reality.
Radial flows are completely neglected in our model, 
though if strong \citep{Schmidt+16}, 
they can have a significant effect on the magnetic pitch angle \citep{Moss+00}.
Finally, certain aspects of dynamo theory remain controversial and poorly understood.
For example, the nature of the magnetic helicity flux is still unclear,
and there can be such effects in addition to the advective and diffusive fluxes 
included in our dynamo model \citep{Subramanian+Brandenburg06,Vishniac12b,Ebrahimi+Bhattacharjee14}.
The need in our fiducial model to slightly enhance 
the dynamo action to attain positive local growth rates hints that
the dynamo mechanism is stronger than implied by the standard estimates.
Another poorly understood effect is the possible modification of the turbulent 
magnetic diffusivity by magnetic field \citep{Brandenburg+08b,Karak+14,Bendre+15,Simard+16}.

\section{Conclusions}
\label{sec:conclusions}
We have used a simple, mean-field dynamo model 
to calculate the strength and direction of the axisymmetric large-scale (or regular) magnetic field 
in nearby disk galaxies for which sufficient data are available.
We obtain a fairly reasonable level of overall agreement between model solutions and observations 
for the magnetic pitch angle $p=\arctan(\mbr/\mbp)$ 
for the galaxies M31, M51, M81, NGC~253, NGC~1566, NGC~6946, and IC~342.
For M33, large pitch angle magnitudes comparable with observed values are obtained,
but detailed agreement is lacking.
For the two barred galaxies in the sample, NGC~1097 and NGC~1365, 
our model does worse,
which is expected as their gas flows and magnetic fields deviate strongly from axial symmetry.
However, agreement for these galaxies is better at larger galactocentric distance,
where effects of the bar are less important.

For the regular magnetic field strength, 
our model agrees with the data to within a factor of a few.
The field strength is less precisely determined than the magnetic pitch angle,
both observationally and theoretically.

We confirm that it is possible to explain large pitch angle magnitudes in galaxies 
by appealing to parameter values different from standard (solar neighborhood) estimates.
Specifically, large pitch angles require a relatively small ratio of disk semi-thickness to turbulent speed $h/u$ 
and/or a relatively large correlation time $\tau$, 
compared to standard estimates of $\sim40$--$50\Myr$ and $10\Myr$, respectively.
 
To obtain locally growing solutions for the mean magnetic field in the kinematic dynamo regime 
for all radii in all galaxies in the sample,
we require that the standard estimate for the kinetic $\alpha$ effect be enhanced by a factor of about $2$.
This suggests that standard estimates for the $\alpha\kin$ effect may be too small.
 
We obtain strong evidence that the ratio $h/u$ tends to increase with galactocentric radius within galaxies.
Assuming, as we have done, that $u$ is approximately constant with radius,
this result implies that the ionized disk within which dynamo action occurs is flared.

Our model produces good agreement with observation 
when we choose the turbulent correlation time to lie somewhere within the range $\tau\sim10$--$20\Myr$,
which is consistent with standard estimates.
The specific value of $\tau$ is, however, likely to vary from galaxy to galaxy.

Extending the present work will require better and more homogeneous magnetic field 
and kinematics data for nearby galaxies,
as well as refinements to the dynamo model to incorporate additional physical effects.
Further, a detailed investigation of the possible connections 
between spiral arms and magnetic pitch angles is still needed.
While our main objective was to make progress toward better tests of dynamo theory,
the present work also demonstrates the promise of inverting the problem 
to probe ISM physics using magnetic fields.

\section*{Acknowledgements}
We are grateful to A.~Fletcher, M.~Krause, C.~Carignan, and M.~Mogotsi for help with interpreting various data,
to N.~Chamandy for providing a code for cross-checking probabilities and for discussions about statistics,
and to L.~F.~S.~Rodrigues, J.~Zwart, and R.-J.~Dettmar for useful discussions.
LC and AS would like to thank K.~Subramanian and IUCAA for hosting them, during which time a part of this work was carried out.
We also appreciate helpful feedback from the referee.
AS acknowledges financial support of the Leverhulme Trust Grant RPG-2014-427 and 
STFC Grant ST/N000900/1 (Project 2).

\bibliographystyle{apj}
\bibliography{refs}

\appendix
\section{Magnetic pitch angle in the no-$z$ approximation}\label{MPANZA}

\begin{figure*}                     
  \begin{tabular}{c c c c} 
\includegraphics[width=0.23\textwidth,clip=true,trim= 20 15 15 10]{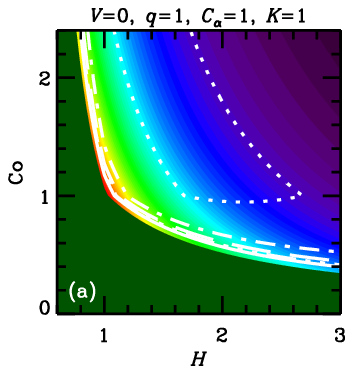}& 
\includegraphics[width=0.23\textwidth,clip=true,trim= 20 15 15 10]{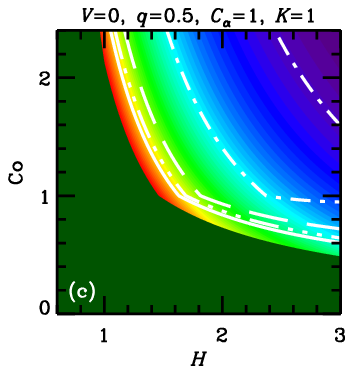}& 
\includegraphics[width=0.23\textwidth,clip=true,trim= 20 15 15 10]{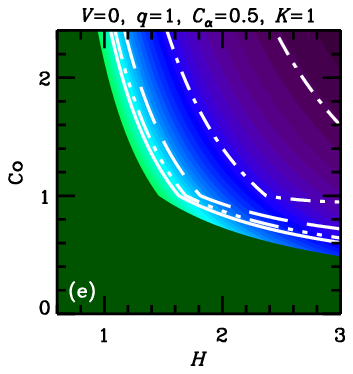}& 
\includegraphics[width=0.23\textwidth,clip=true,trim= 20 15 15 10]{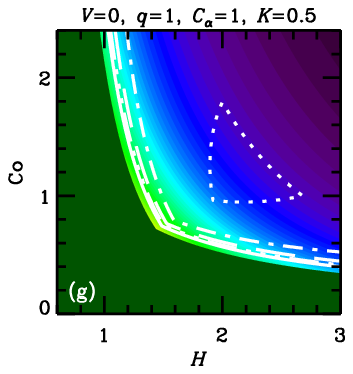}\\
\includegraphics[width=0.23\textwidth,clip=true,trim= 20 15 15 10]{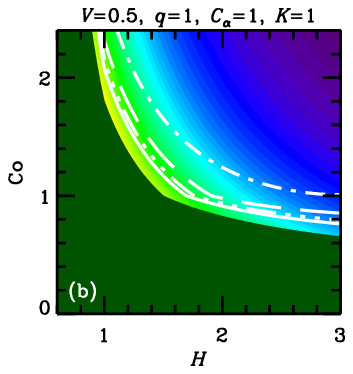}&
\includegraphics[width=0.23\textwidth,clip=true,trim= 20 15 15 10]{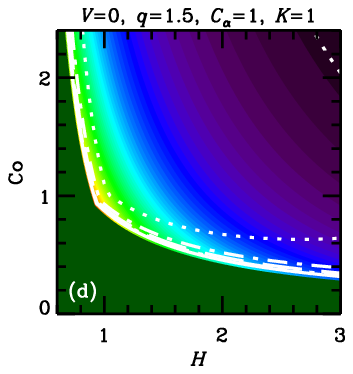}&
\includegraphics[width=0.23\textwidth,clip=true,trim= 20 15 15 10]{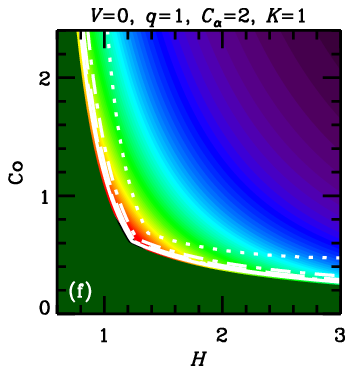}  &
\includegraphics[width=0.23\textwidth,clip=true,trim= 20 15 15 10]{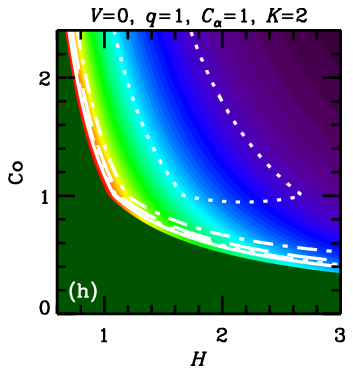}\\
  \end{tabular}
  \begin{center}
    \includegraphics[width=0.7\textwidth,clip=true,trim=-20 20  0 0]{col_p_V0_q1_K1_Calp1_Cbet1.eps} 	
  \end{center}
  \caption{Saturated mean magnetic pitch angle (color) for various sets of parameter values noted at the top of each panel,
           for the analytical solution, as a function of the dimensionless disk semi-thickness $H$ and Coriolis number $\Coriolis$.
           Contours denote the local $10^3$-folding time in local galactic rotation periods $T_3=6$ (dotted), 
           $12$ (dashed-dotted), $18$ (dashed), $24$ (dash-triple-dotted), and $30$ (solid).
           Panel~(a) shows the solution for fiducial parameter values (no outflow, locally flat rotation curve),
           while panel~(b) includes a strong outflow.
           Each column shows how the solution changes when the value of a single parameter is varied from the fiducial case.
           Dark green denotes the region of parameter space with decaying solutions in the kinematic regime.
           \label{fig:psat_anal}
          }            
\end{figure*} 

\section{Mean magnetic field in the case of negligible outflow}
\label{sec:V0}
For the pitch angle in the saturated state we find
\begin{equation}
  \label{psatV0}
  \tan p\sat|_{V=0}= -\frac{\pi^2\tau}{12q\Omega}\left(\frac{u}{h}\right)^2.
\end{equation}
For the local growth rate we have
\begin{subnumcases}{ \label{gammaV0}
  \gamma|_{V=0}= \sqrt{\frac{2q}{\pi}}\Omega\tau \left(\frac{u}{h}\right)\times}
      \sqrt{C_\alpha} -\Lambda,  \\
      \sqrt{\frac{C_\alpha}{\Omega\tau}} -\Lambda, \\
      \sqrt{\frac{K}{\Omega\tau^2}\left(\frac{h}{u}\right)} -\Lambda,
\end{subnumcases}
with
\begin{equation}
  \label{Lambda}
  \Lambda=\frac{1}{3\Omega}\left(\frac{u}{h}\right)\sqrt{\left(\frac{\pi}{2}\right)^5\frac{1}{q}}.
\end{equation}
For the field strength in the saturated state we have
\begin{equation}
  \label{BsatV0}
  B\sat^2|_{V=0}= \frac{\pi^3\Strouhal^2\Sigma_I R_\kappa \tau^2 u}{\xi}\left(\frac{u}{h}\right)^3
                  \left(\frac{\tan^2p\kin}{\tan^2p\sat|_{V=0}}-1\right),
\end{equation}
where $\tan^2p\kin$ is given by equation~\eqref{pkin} and $\tan^2p\sat|_{V=0}$ by equation~\eqref{psatV0}.


\end{document}